\documentclass[journal]{IEEEtran}
\IEEEoverridecommandlockouts
\usepackage{cite}
\usepackage{graphicx}
\usepackage{amsmath}
\usepackage{amssymb}
\usepackage{algorithmicx}
\usepackage{algorithm}
\usepackage{color}
\usepackage{pdfcomment}

\allowdisplaybreaks

\begin{document}

\title{MIMO Transmission through\\ Reconfigurable Intelligent Surface:\\ System Design, Analysis, and Implementation}

\author{
\IEEEauthorblockN{Wankai Tang, Jun Yan Dai, Ming Zheng Chen, Kai-Kit Wong,\\ Xiao Li, Xinsheng Zhao, Shi Jin, Qiang Cheng, and Tie Jun Cui}

\thanks{This article was presented in part at the IEEE SPAWC 2020.}
\thanks{W. Tang, X. Li, X. Zhao, and S. Jin are with the National Mobile Communications Research Laboratory, Southeast University, Nanjing, China (e-mail: tangwk@seu.edu.cn, li\_xiao@seu.edu.cn, xszhao@seu.edu.cn, and jinshi@seu.edu.cn).}
\thanks{J. Y. Dai, M. Z. Chen, Q. Cheng, and T. J. Cui are with the State Key Laboratory of Millimeter Waves, Southeast University, Nanjing, China (e-mail: junyandai@seu.edu.cn, mingzhengc@seu.edu.cn, qiangcheng@seu.edu.cn, and tjcui@seu.edu.cn).}
\thanks{K.-K. Wong is with the Department of Electronic and Electrical Engineering, University College London, London, United Kingdom (e-mail: kai-kit.wong@ucl.ac.uk).}
}

\maketitle

\begin{abstract}

Reconfigurable intelligent surface (RIS) is a new paradigm that has great potential to achieve cost-effective, energy-efficient information modulation for wireless transmission, by the ability to change the reflection coefficients of the unit cells of a programmable metasurface. Nevertheless, the electromagnetic responses of the RISs are usually only phase-adjustable, which considerably limits the achievable rate of RIS-based transmitters. In this paper, we propose an RIS architecture to achieve amplitude-and-phase-varying modulation, which facilitates the design of multiple-input multiple-output (MIMO) quadrature amplitude modulation (QAM) transmission. The hardware constraints of the RIS and their impacts on the system design are discussed and analyzed. Furthermore, the proposed approach is evaluated using our prototype which implements the RIS-based MIMO-QAM transmission over the air in real time.

\end{abstract}


\section{Introduction}
The fifth-generation (5G) mobile communications is being rolled out around the world, as fully digital massive multiple-input multiple-output (MIMO) antennas at the base stations (BSs) are tasked to provide the technological leaps that are being sought in 5G \cite{HW}. The use of millimeter wave (mmWave) adds another dimension to address the spectrum shortage problem \cite{mmwave}. Together with a number of other innovative technologies, 5G is aimed to handle demanding applications such as virtual/augmented reality (VAR), holographic projection, autonomous driving, and tactile Internet, to name just a few, but provision of such applications in full scale is not expected until the sixth-generation (6G) that keeps researchers working in the next decade \cite{6G}.

Looking ahead to what technologies may deliver 6G, extending the spectrum to the terahertz (THz) band \cite{terahertz} and upscaling massive MIMO (resulting in ultra-massive MIMO (UM-MIMO) \cite{ummimo}), appear to be a natural next step. Further, large intelligent surface (LIS) \cite{activeLIS} and holographic MIMO \cite{holomimo} also emerge as some 6G enabling technologies to obtain extraordinary spatial diversity for much improved performance. However, the very high operating frequency at the terahertz and the extremely large number of radio-frequency (RF) chains required by UM-MIMO technologies will lead to very high hardware implementation costs and excessive energy consumption.

Against this background, reconfigurable intelligent surface (RIS)\footnote{Also known as intelligent reflecting surface (IRS) and passive large intelligent surface (LIS) in the literature, which are often expected to have advanced reconfigurable reflecting elements on a surface.} \cite{RIS1,RIS2,RIS3,RIS4,RIS5} is a new paradigm that can flexibly manipulate electromagnetic (EM) waves, and provide a hardware architecture that radically alleviates the implementation issues. An RIS is made of a programmable metasurface \cite{Meta1,Meta2,Meta3}, which can be controlled by external signals to realize real-time manipulation on the EM responses of the reflected waves, such as the phase and amplitude. The programmable EM properties of the RISs empower them the ability to engineer radio signals that are appealing for wireless communications. In particular, the RIS-based wireless transmitter can directly perform modulation on the EM carrier signals, without the need for conventional RF chains, thereby having great potential for realizing UM-MIMO and holographic MIMO technologies~\cite{ELfeature}.
\vspace{-0.1cm}
\subsection{Related Work}\label{RelatedWork}
There have been attempts to achieve cost-effective wireless transmitters using novel hardware architectures. In these studies, it was remarkably reported that only the single-tone carrier signal needs to be amplified by the power amplifier (PA), and the baseband data is directly modulated onto the carrier signal. In particular, the direct antenna modulation (DAM) technology was proposed in \cite{DAM1,DAM2,DAM3} to directly generate modulated RF signals through the time-varing antennas with positive-intrinsic-negative (PIN) diodes. Nevertheless, DAM only supports the inefficient on-off keying (OOK) modulation scheme. In \cite{DPSM1} and \cite{DPSM2}, an antenna array with elements driven by the phase shifters and carrier signals was explored to realize direct phase modulation, but it suffers from low data rate due to the slow update rate of the phase shifters.

Of particular attention was the several RIS-based transmitter architectures that were investigated, e.g., \cite{RIS-based1,RIS-based2,RIS-based3,RIS-based4,RIS-based5,RIS-based6,RIS-based7,RIS-based8,RIS-based9}. In \cite{RIS-based1}, an RIS-based binary frequency shift-keying (BFSK) transmitter with a simplified architecture was proposed. Subsequently, in \cite{RIS-based2} and \cite{RIS-based3}, the experiments of RIS-based quadrature phase shift keying (QPSK) transmission over the air were demonstrated. Elaborately designed RISs that achieved a $360^\circ$ phase shift coverage were adopted to develop an 8-phase shift keying (8PSK) wireless communication prototype in \cite{RIS-based4} and \cite{RIS-based5}, respectively. Recently, \cite{RIS-based6} further realized multi-modulation schemes using an RIS for wireless communications. Moreover, a mathematical framework by using probabilistic tools was provided in \cite{RIS-based7} and \cite{RIS-based8} to evaluate the theoretical symbol error probability (SEP) of an RIS-based transmitter. In addition, an RIS-based modulation and resource allocation scheme without causing interference with existing users was proposed to enhance the achievable system sum-rate~\cite{RIS-based9}.

The RIS-based transmitters have shown great potential to bring a new paradigm that naturally integrates between signal processing algorithms in information science and hardware resources made of programmable metamaterial unit cells for future-generation wireless communications. However, the limitation of existing research on RIS-based transmitters appears to be the lack of an analytical formulation that describes the system model considering the physics and EM nature of the RISs. Moreover, the hardware constraints of the RISs, such as phase dependent amplitude and discrete phase shift{\footnote {The impact of discrete phase shift has been taken into account in the literature\cite{RIS5,RIS-discrete1,RIS-discrete2}, in which RIS works as a reflector for passive beamforming, rather than  the RIS-based transmitter presented in this paper.}}, have been largely ignored in most existing works. Additionally, prior prototypes on RIS-based wireless transmitters were all limited to basic single-input single-output (SISO) communications. Whether RIS-based MIMO transmission is possible is not understood, and it is also not clear if the benefits of RIS still prevail when using it for realizing UM-MIMO or holographic MIMO wireless communications even if it is possible.
\vspace{-0.1cm}
\subsection{Main Contributions}\label{MainContributions}
This paper aims to investigate the feasibility of using RIS for MIMO wireless transmission for higher-order modulation by presenting an analytical modelling of the RIS-based system and providing experimental results from a prototype which has been built. The main contributions of this paper are summarized as follows:
\begin{enumerate}
\item We present a mathematical model that characterizes RIS-based MIMO transmission considering the physics and EM nature of the RISs. The system model reveals that the working principle and basic expression of an RIS-based MIMO wireless communication system is the same as that of the conventional non-RIS based system.
\item In addition, we introduce a non-linear modulation technique to realize high-order modulation under the constant envelope constraint, and apply it in the RIS-based MIMO transmission. Furthermore, the hardware constraints of the RISs including phase dependent amplitude and discrete phase shift, and their impacts on the RIS-based MIMO quadrature amplitude modulation (QAM) wireless communication system design are discussed and analyzed.
\item By using the proposed architecture, we present the world's first prototype that implements real-time RIS-based MIMO-QAM wireless communication. In our prototype, a varactor-diode-based programmable metasurface is utilized. The power consumption of the metasurface and its control circuit board is about $0.7$W, and the achievable data rate of the prototype system is $20$ Mbps. The experimental results validate that the proposed RIS-based MIMO-QAM wireless transmitter architecture is robust, and potentially a cost-effective and energy-efficient hardware architecture for emerging wireless communication systems with an extremely large aperture, such as UM-MIMO and holographic MIMO.
\end{enumerate}
\vspace{-0.1cm}
\subsection{Organization}\label{PaperOrganization}
The remainder of this paper is organized as follows. Section II introduces the fundamentals of a basic RIS-based transmitter, and then develops the system model of an RIS-based MIMO wireless communications. Section III presents a method of achieving MIMO-QAM transmission and beamforming through RIS. In Section IV, we discuss the hardware constraints of RISs, and provide our transceiver design of an RIS-based 2$\times$2 MIMO-QAM wireless communication system. The implementation and experimental results of the prototype for RIS-based 2$\times$2-MIMO 16QAM transmission are presented in Section V. Section VI concludes the paper.

\section{System Model}
This section reviews the fundamentals of RIS-based transmitter and develops its system model, which will be used in the following sections for system design, analysis and implementation.
\vspace{-0.1cm}
\subsection{Fundamentals of RIS-based Transmitter}\label{FundamentalTransmitter}
\subsubsection{RIS-based Modulation}\label{RIS-basedModulation}
\begin{figure}
\centering
\includegraphics[scale = 0.55]{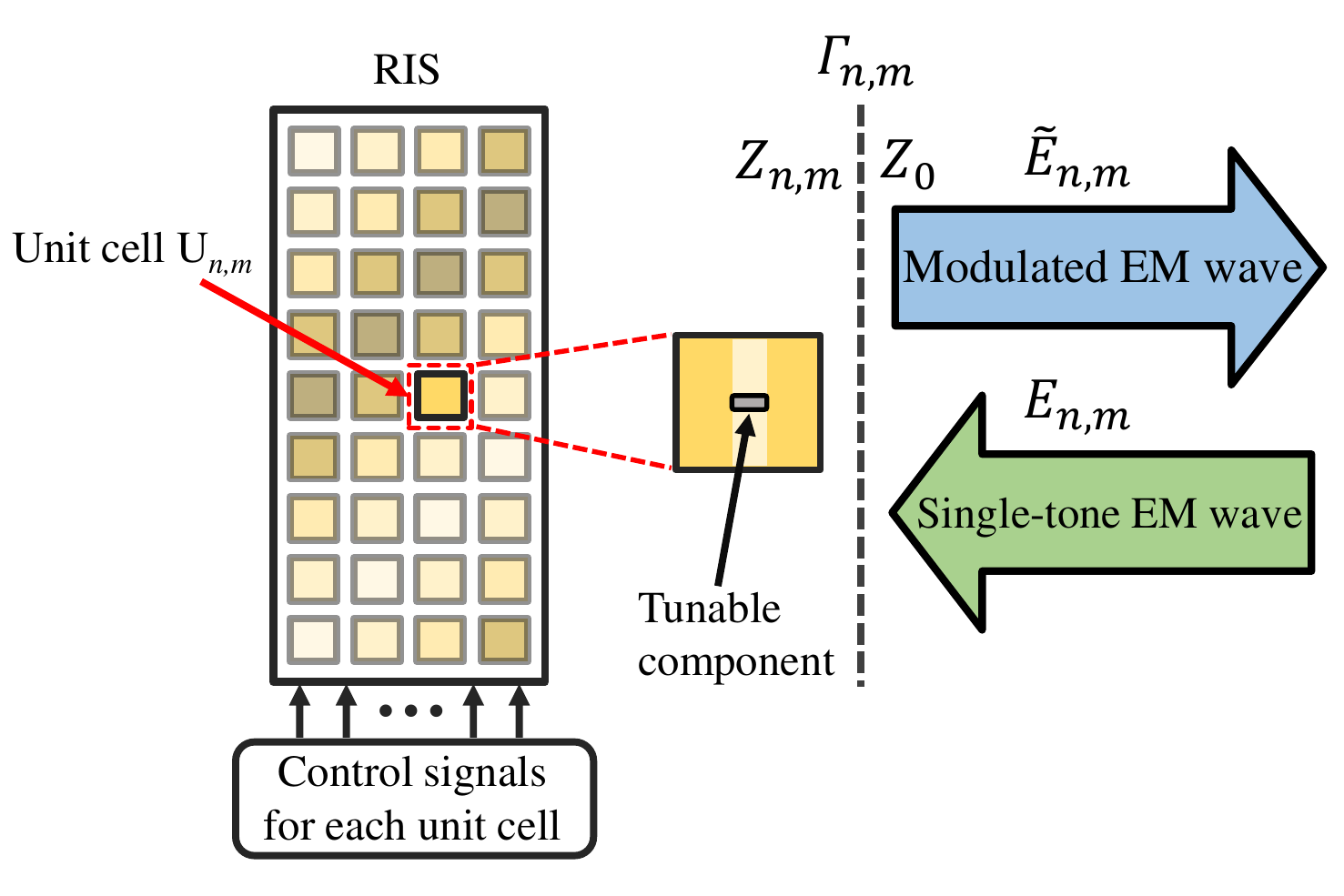}
\caption{Illustration of the RIS-based modulation.}
\label{diagramRISmodulation}
\vspace{-0.45cm}
\end{figure}

As an emerging technology that can flexibly manipulate EM waves, technically speaking, RIS is indeed a programmable metasurface composed of sub-wavelength unit cells within the range of $\frac{\lambda}{10}$ and $\frac{\lambda}{2}$. As shown in Fig. \ref{diagramRISmodulation}, the unit cells of the RIS are regularly arranged and thus form a two-dimensional artificial structure. The unit cell with tunable EM properties is typically comprised of elaborately designed metal pattern, dielectric and tunable component. The external control signal of each unit cell can change the electrical parameters of the tunable component, thereby altering the EM responses of the unit cell, such as the phase and amplitude. Taking the unit cell $U_{n,m}$ in the $n^{th}$ row and $m^{th}$ column as an example, let $E_{n,m}$, $\widetilde E_{n,m}$, $Z_0$, $Z_{n,m}$ and $\varGamma _{n,m}$ represent the EM wave impinging on $U_{n,m}$, the EM wave reflected from $U_{n,m}$, the characteristic impedance of the air, the equivalent load impedance of $U_{n,m}$, and the reflection coefficient of $U_{n,m}$, respectively. The reflection coefficient is a parameter that describes the complex-valued fraction of the EM wave reflected by an impedance discontinuity in the transmission medium, which can be expressed as
\begin{equation}\label{ss1}
{\varGamma _{n,m}} = {A_{n,m}}{e^{j{\varphi _{n,m}}}},
\end{equation}
where ${A_{n,m}}$ and ${\varphi _{n,m}}$ represent the controllable amplitude and phase shift of $U_{n,m}$, respectively. According to the definition of the reflection coefficient, we have
\begin{equation}\label{ss2}
{\widetilde E_{n,m}} = {\varGamma _{n,m}}{E_{n,m}} = {A_{n,m}}{e^{j{\varphi _{n,m}}}}{E_{n,m}}.
\end{equation}
In addition, the reflection coefficient $\varGamma _{n,m}$ of the unit cell $U_{n,m}$ is determined by its equivalent load impedance $Z_{n,m}$ and the impedance towards the source $Z_0$, which is written as\cite{Book0}
\begin{equation}\label{ss3}
{\varGamma _{n,m}} = \frac{{{Z_{n,m}} - {Z_0}}}{{{Z_{n,m}} + {Z_0}}}.
\end{equation}
By combining (\ref{ss1}) and (\ref{ss3}), the amplitude and the phase of the reflection coefficient $\varGamma _{n,m}$ can be obtained as
\begin{equation}\label{ss4}
{A_{n,m}} = \left| {\frac{{{Z_{n,m}} - {Z_0}}}{{{Z_{n,m}} + {Z_0}}}} \right|,
\end{equation}
and
\begin{equation}\label{ss5}
{\varphi _{n,m}} = \arctan \left( {\frac{{\operatorname{Im} \left( {\frac{{{Z_{n,m}} - {Z_0}}}{{{Z_{n,m}} + {Z_0}}}} \right)}}{{\operatorname{Re} \left( {\frac{{{Z_{n,m}} - {Z_0}}}{{{Z_{n,m}} + {Z_0}}}} \right)}}} \right).
\end{equation}

As the equivalent load impedance $Z_{n,m}$ can be adjusted by the external control signal, (\ref{ss4}) and (\ref{ss5}) reveal the reflection amplitude and phase altering principle of the unit cells of the RIS. In particular, when the RIS is employed as the wireless transmitter, the incident EM wave in (\ref{ss2}) is a single-tone EM wave with frequency $f_c$ and amplitude $A_c$, which acts as the carrier signal. Then  (\ref{ss2}) can be further expressed as
\begin{equation}\label{ss6}
{\widetilde E_{n,m}} = {A_{n,m}}{e^{j{\varphi _{n,m}}}}{A_c}{e^{j2\pi {f_c}t}} = {A_c}{A_{n,m}}{e^{j\left( {2\pi {f_c}t + {\varphi _{n,m}}} \right)}},
\end{equation}
which indicates that the adjustable ${A_{n,m}}$ and ${\varphi _{n,m}}$ can achieve amplitude modulation and phase modulation on the air-fed carrier signal, which is referred to as \textbf{\emph{RIS-based modulation}}. If all the unit cells of the RIS are controlled by the same external control signal, then the entire RIS will perform the same modulation on the air-fed carrier signal, thereby realizing the basic SISO wireless communications.

\subsubsection{RIS-based Multi-channel Transmitter}\label{RIS-basedMultiTransmitter}
\begin{figure}
\centering
\includegraphics[scale = 0.521]{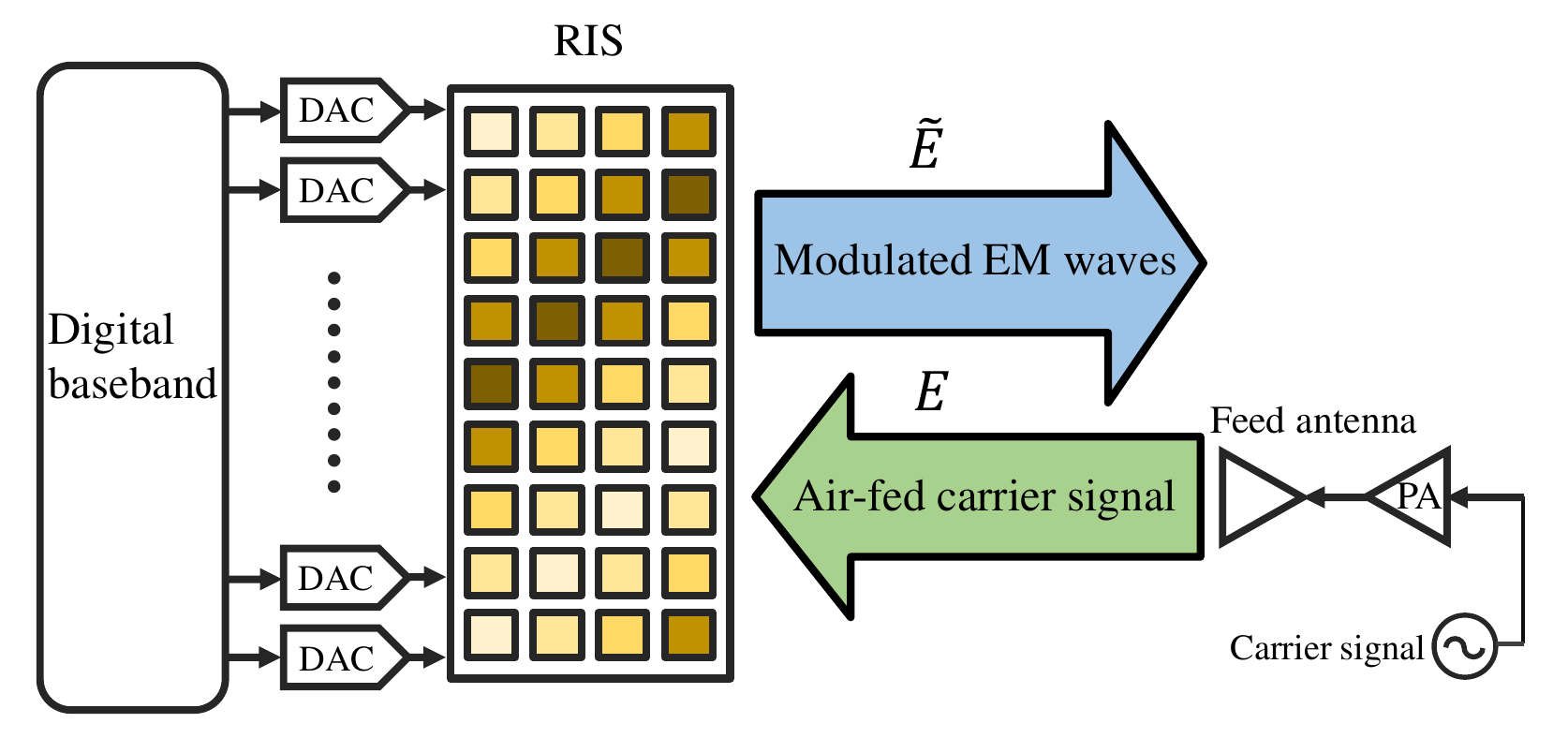}
\caption{An RIS-based multi-channel transmitter.}
\label{diagramRIStransmission}
\vspace{-0.5cm}
\end{figure}

Since the reflection coefficient of each unit cell of an RIS can be controlled independently by a dedicated control signal, an RIS-based transmitter can achieve multi-channel transmission. Fig. \ref{diagramRIStransmission} depicts the diagram of an RIS-based multi-channel transmitter. As shown in Fig. \ref{diagramRIStransmission}, the digital baseband contains multiple bitstreams, which can be mapped to the control signals of the unit cells through the digital-to-analog converters (DACs). The maximum number of the bitstreams that can be transmitted simultaneously is the same as the number of unit cells of the RIS, that is, each unit cell is controlled by one dedicated DAC in this case. The total reflected EM wave observed at a certain location is the superposition of the reflected EM waves from all the unit cells of the RIS.

In the RIS-based transmitter, the baseband modules are directly connected to the radiating elements (unit cells) without the need for conventional RF chains, and therefore the RIS-based transmitter is considered an \textbf{\emph{RF chain-free transmitter}}. In addition, the RIS-based transmitter only requires one narrowband power amplifier (PA) to control the power of the air-fed carrier signal, thus circumventing the nonlinearity issue of PAs. Compared with the conventional multi-channel wireless transmitter, these characteristics of RIS-based transmitter significantly reduce the hardware cost and complexity, rendering it especially attractive for the emerging ultra-massive MIMO (UM-MIMO) and holographic MIMO wireless communication technologies.
\vspace{-0.1cm}
\subsection{Communication Model}\label{SystemModel}
\newcommand{\tabincell}[2]{\begin{tabular}{@{}#1@{}}#2\end{tabular}}  
\begin{table*}
\centering
\footnotesize
\caption{Notations and definitions for RIS-based MIMO wireless communication system.}\label{Notations}
\begin{tabular}{|c|l|c|l|}
\hline
\textbf{Notation} & \textbf{Definition}  & \textbf{Notation} & \textbf{Definition}\\
\hline
$N$ &  Number of rows of unit cells & $U_{n,m}$ &  Unit cell in the $n^{th}$ row and $m^{th}$ column\\
\hline
$M$ &  Number of columns of unit cells & $\lambda$ & Wavelength of the signal\\
\hline
$\varGamma _{n,m}$ &  Reflection coefficient of $U_{n,m}$& $f_c$ & Frequency of the carrier signal\\
\hline
$A_{n,m}$ &  Amplitude component of $\varGamma _{n,m}$ & $S$ &  Incident energy flux density on each unit cell\\
\hline
${\varphi _{n,m}}$ &  Phase component of $\varGamma _{n,m}$ & $d_{n,m}^k$ & Distance between $U_{n,m}$ and $k^{th}$ receiving antenna\\
\hline
$d_x$ &  Width of each unit cell & $(\theta _{n,m}^{AOD,k},\phi _{n,m}^{AOD,k})$ & \tabincell{l}{Angle of departure (AoD) of the signal between \\$U_{n,m}$ and $k^{th}$ receiving antenna}\\
\hline
$d_y$ &  Length of each unit cell & $(\theta _{n,m}^{AOA,k},\phi _{n,m}^{AOA,k})$ & \tabincell{l}{Angle of arrival (AoA) of the signal between \\$U_{n,m}$ and $k^{th}$ receiving antenna}\\
\hline
$G$ &  Gain of each unit cell & $G_r$ & Gain of each receiving antenna\\
\hline
$F\left( {\theta ,\varphi } \right)$ & \tabincell{l}{Normalized power radiation\\pattern of each unit cell}& ${F^{rx}}(\theta,\varphi)$ & \tabincell{l}{Normalized power radiation pattern of each \\receiving antenna}\\
\hline
\end{tabular}
\end{table*}

\begin{figure*}
\centering
\includegraphics[scale = 0.6]{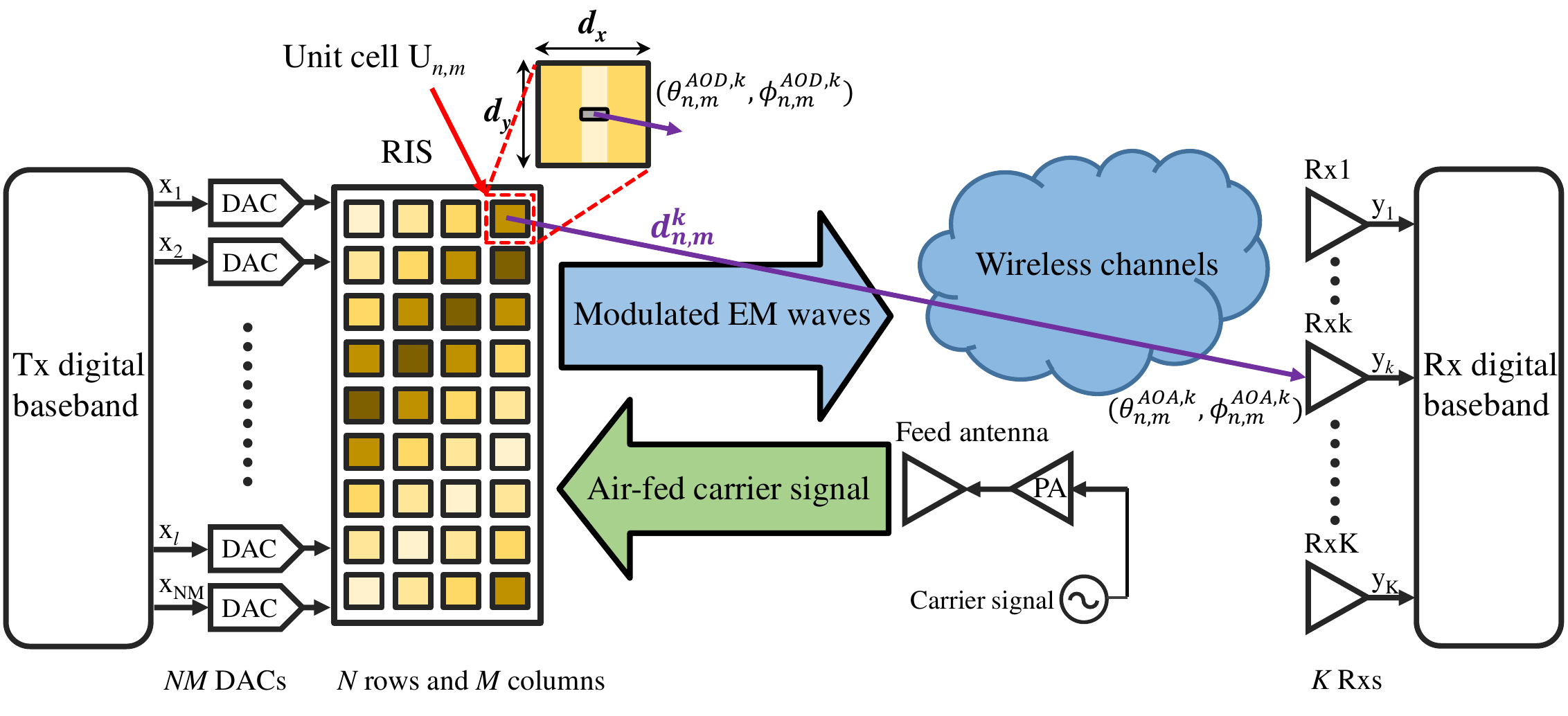}
\caption{An RIS-based MIMO wireless communication system.}
\label{RISMIMOcommunication}
\vspace{-0.2cm}
\end{figure*}
We consider a general RIS-based MIMO wireless communication system as shown in Fig. \ref{RISMIMOcommunication}, in which each unit cell is controlled by a dedicated DAC. The RIS consists of $N$ rows and $M$ columns of unit cells. The unit cell in the $n^{th}$ row and $m^{th}$ column, $U_{n,m}$, has the reflection coefficient $\varGamma _{n,m}$, for $n \in \left[ {1,N} \right]$ and $m \in \left[ {1,M} \right]$. Let $d_x$, $d_y$, $G$, and $F\left( {\theta ,\varphi } \right)$ denote the width, length, gain, and the normalized power radiation pattern of $U_{n,m}$, respectively\cite{Book1,Pathloss}. Assume that the air-fed carrier signal (incident EM wave) with frequency $f_c$ is a uniform plane wave perpendicular to the RIS{\footnote{We assume that the RIS is uniformly illuminated by the feed antenna throughout this paper, i.e., the amplitude and phase of the incident signal on each unit cell of the RIS are consistent. This assumption is reasonable in practice, because the feed antenna can be placed directly in front of the RIS to ensure normal incidence under far-field condition, and there is no obstacle between the RIS and the feed antenna.}} and its energy flux density on $U_{n,m}$ is denoted as $S$. Also, it is assumed that there are $K$ antennas at the receiver side. Let $d_{n,m}^k$, $\theta _{n,m}^{AOD,k}$, $\phi _{n,m}^{AOD,k}$, $\theta _{n,m}^{AOA,k}$ and $\phi _{n,m}^{AOA,k}$ represent the distance between $U_{n,m}$ and the $k^{th}$ receiving antenna, the elevation angle and the azimuth angle from $U_{n,m}$ to the $k^{th}$ receiving antenna, the elevation angle and the azimuth angle from the $k^{th}$ receiving antenna to $U_{n,m}$, respectively. That is, $(\theta _{n,m}^{AOD,k},\phi _{n,m}^{AOD,k})$ is the angle of departure (AoD) and $(\theta _{n,m}^{AOA,k},\phi _{n,m}^{AOA,k})$ is the angle of arrival (AoA) of the signal between $U_{n,m}$ and the $k^{th}$ receiving antenna, for $k \in \left[ {1,K} \right]$. The receiving antennas all have the same antenna design with a normalized power radiation pattern ${F^{rx}}(\theta,\varphi)$ and antenna gain $G_r$. All the above notations and definitions are summarized in Table \ref{Notations}.

The following result presents the received signal of the $k^{th}$ receiving antenna in the above RIS-based MIMO wireless communication system in the case of free-space propagation.

{\textbf{Theorem 1.}} For free-space propagation without considering noise, the received signal of the $k^{th}$ receiving antenna in the RIS-based MIMO wireless communication system is given as (\ref{ss7}) shown at the top of the next page.

\begin{figure*}[!t]
\begin{align}\label{ss7}
{y_k} = \sum\limits_{m = 1}^M \sum\limits_{n = 1}^N \frac{{\sqrt {G{G_r}{\lambda ^2}F\left( {\theta _{n,m}^{AOD,k},\phi _{n,m}^{AOD,k}} \right){F^{rx}}\left( {\theta _{n,m}^{AOA,k},\phi _{n,m}^{AOA,k}} \right)S{d_x}{d_y}} }}{{4\pi d_{n,m}^k}}{e^{\frac{{ - j2\pi d_{n,m}^k}}{\lambda }}}{\varGamma _{n,m}}{e^{j2\pi {f_c}t}}
\end{align}
\hrulefill
\end{figure*}

\emph{Proof}: See Appendix A. \hfill $\blacksquare$

Theorem 1 reveals that the received signal of the $k^{th}$ receiving antenna is the superposition of the signal reflected by all the unit cells towards it. For the transmission path from each unit cell to the $k^{th}$ receiving antenna, the amplitude of the received signal is proportional to the square root of the gains of the unit cell and the receiving antenna, the wavelength, the square root of the normalized power radiation patterns of the unit cell and the receiving antenna, the square root of the incident energy flux density, and the square root of the size of the unit cell. In addition, the amplitude is inversely proportional to the distance between the unit cell and the receiving antenna. Furthermore, the phase of the received signal in each path is related to the phase shift caused by the transmission distance and the phase shift induced by the reflection coefficient.

Theorem 1 gives the expression of the received signal of the $k^{th}$ receiving antenna in the free-space propagation case in the absence of noise. In this particular case, the wireless channel between the unit cell $U_{n,m}$ and the $k^{th}$ receiving antenna can be expressed as
\begin{equation}\label{ss8}
\begin{aligned}
&h_{n,m}^{k,freespace}=\\
&\frac{{\sqrt {G{G_r}{\lambda ^2}F\left( {\theta _{n,m}^{AOD,k},\phi _{n,m}^{AOD,k}} \right){F^{rx}}\left( {\theta _{n,m}^{AOA,k},\phi _{n,m}^{AOA,k}} \right)} }}{{4\pi d_{n,m}^k}}\\
&\times {e^{\frac{{ - j2\pi d_{n,m}^k}}{\lambda }}}.
\end{aligned}
\end{equation}
As a consequence, (\ref{ss7}) can be rewritten as
\begin{equation}\label{ss9}
{y_k} = \sum\limits_{m = 1}^M {\sum\limits_{n = 1}^N {h_{n,m}^{k,freespace}\sqrt p {A_{n,m}}{e^{j{\varphi _{n,m}}}}{e^{j2\pi {f_c}t}}} },
\end{equation}
in which $p = S{d_x}{d_y}$ represents the transmission power of each unit cell. The expression (\ref{ss9}) describes the basic principle of RIS-based MIMO wireless communication, with programmable amplitude ${A_{n,m}}$ and phase shift $\varphi _{n,m}$ of the unit cells.

Now, we extend the model to the flat-fading case. Denoting the channel between the unit cell $U_{n,m}$ and the $k^{th}$ receiving antenna as $h_{n,m}^k$, and the noise at the $k^{th}$ receiving antenna as $n_k$, we can write the channel between the RIS and the $k^{th}$ receiving antenna as
\begin{equation}\label{ss11}
{{\bf{h}}_k}{=}\left[ {h_{1,1}^k,h_{1,2}^k, \dots ,h_{1,M}^k,h_{2,1}^k, \dots ,h_{N,M}^k} \right]\in {\mathbb{C}^{1 \times NM}}.
\end{equation}
Furthermore, we denote the transmitted baseband symbols in vector as
\begin{equation}\label{ss12}
\begin{aligned}
{\bf{x}} =& \left[ {{x_1},{x_2}, \dots ,{x_M},{x_{M + 1}}, \dots ,{x_{NM}}} \right]^T\\
 =& \left[ {A_{1,1}}{e^{j{\varphi _{1,1}}}},{A_{1,2}}{e^{j{\varphi _{1,2}}}}, \dots ,{A_{1,M}}{e^{j{\varphi _{1,M}}}},\right.\\
 &\left.{A_{2,1}}{e^{j{\varphi _{2,1}}}}, \dots ,{A_{N,M}}{e^{j{\varphi _{N,M}}}} \right]^T \in {\mathbb{C}^{NM \times 1}}.
\end{aligned}
\end{equation}
As a result, the baseband expression of the received signal of the $k^{th}$ receiving antenna can be written as
\begin{equation}\label{ss10}
{y_k}{=}\sum\limits_{m = 1}^M {\sum\limits_{n = 1}^N {h_{n,m}^k\sqrt p {A_{n,m}}{e^{j{\varphi _{n,m}}}}{+}{n_k}} }  =\sqrt p {{\bf{h}}_k}{\bf{x}} + {n_k}.
\end{equation}
Based on (\ref{ss10}), the baseband expression of the RIS-based MIMO wireless communication system can be written as
\begin{equation}\label{ss13}
{\mathbf{y}} = \sqrt p {\mathbf{Hx}} + {\mathbf{n}},
\end{equation}
where ${\mathbf{y}}= {\left[ {{y_1},\dots,{y_K}} \right]^T} \in {\mathbb{C}^{K \times 1}}$ is the received signal vector at the receiver side as shown in Fig. \ref{RISMIMOcommunication}, ${\mathbf{H}}= {\left[ {{{\mathbf{h}}_1},\dots,{{\mathbf{h}}_K}} \right]^T}\in {\mathbb{C}^{K \times NM}}$ denotes the wireless channel matrix between the RIS and the receiver, and ${\mathbf{n}} \in {\mathbb{C}^{K \times 1}}$ is the noise vector at the receiver side.

On one hand, (\ref{ss13}) reveals that the essential principle of RIS-based MIMO wireless communication system is the same as that of the conventional one\cite{Book2}. That is, the multi-channel transmitter modulates the carrier signal and then radiates the modulated RF signals to the multi-channel receiver over the wireless channels. The difference is that every channel of the conventional transmitter modulates the carrier signal by the in-phase and quadrature (IQ) signals and radiates the modulated RF signal through an antenna element, while every channel of the RIS-based RF chain-free transmitter modulates the air-fed carrier signal by the reflection coefficients of the unit cells and radiates the modulated RF signal through these unit cells. The RIS-based transmitter shown in Fig. \ref{RISMIMOcommunication} only differs in terms of the hardware architecture, while sharing the same essential principle and basic mathematical expression with the conventional transmitter. Therefore, the proposed RIS-based MIMO transmission has the same diversity and coding gains with the conventional MIMO systems, which are related to the specific transmission scheme. Moreover, the extensively studied MIMO transmission schemes and algorithms can be applied in the RIS-based MIMO wireless communication system. On the other hand, the unique advantages of RIS-based MIMO for being chain-free and power-efficient make it an attractive architecture for emerging wireless communication systems, such as UM-MIMO and holographic MIMO that conventional transmitters are hard to realize due to the hardware cost and heat dissipation issues.

\section{RIS-based MIMO-QAM Transmission}
\begin{figure}
\centering
\includegraphics[scale = 0.625]{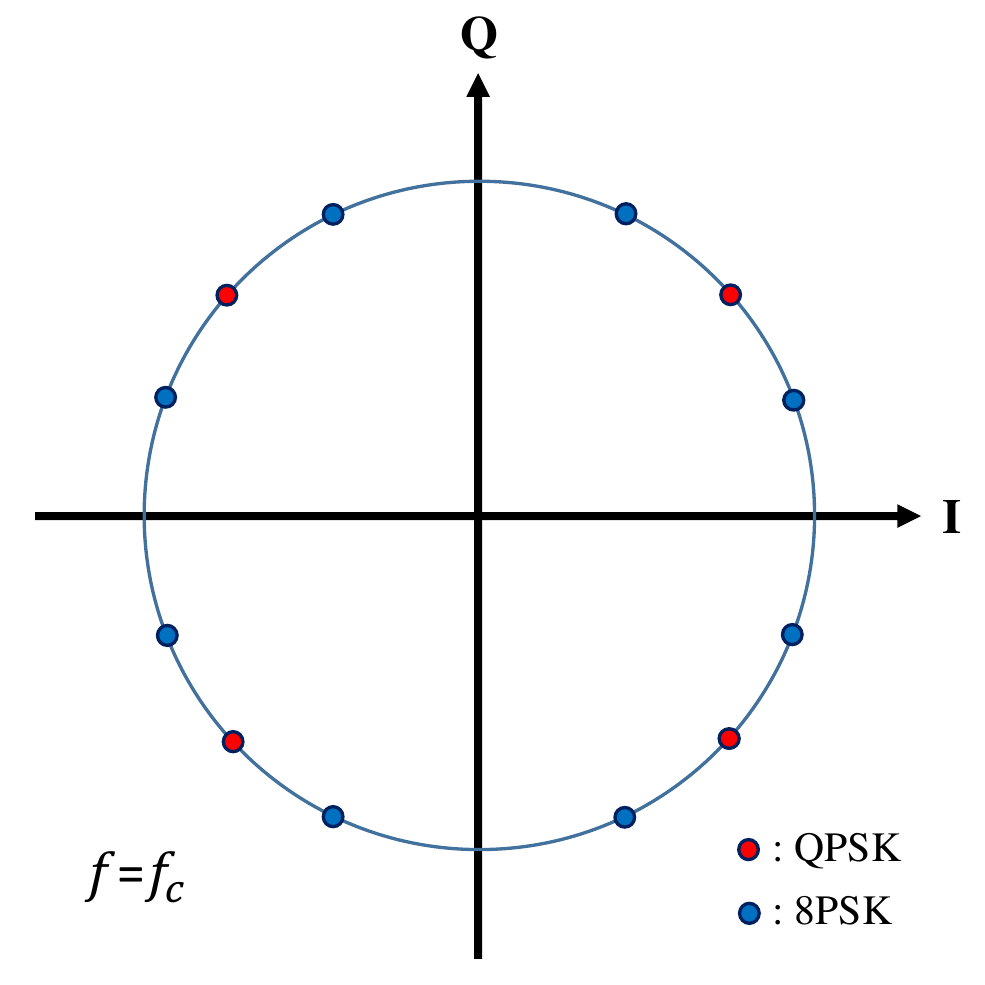}
\caption{Illustration of RIS-based constant envelope modulations.}
\label{RISCEmodulation}
\vspace{-0.2cm}
\end{figure}

In this section, we present our design for RIS-based QAM modulation and MIMO transmission. In conventional wireless transmitters, every RF chain has two independent baseband signals to modulate the carrier signal, i.e., the IQ components, thus realizing independent modulation of the amplitude and phase of the carrier signal. Therefore, QAM can be naturally achieved through the conventional wireless transmitters. However, QAM is hard to achieve using RIS-based transmitters. The previous prototyping works of RIS-based transmitter mainly focused on constant envelope modulations, such as QPSK and 8PSK, as shown in Fig. \ref{RISCEmodulation}. This is because the reflection amplitude and phase responses of a unit cell are usually strongly coupled. Each unit cell of most RISs is designed to have only one external signal to control its equivalent load impedance $Z_{n,m}$, i.e., there is only one control degree of freedom for the EM response of the unit cell. There have been unit cell designs with two or more control signals, but multiple control signals of each unit cell are usually for manipulation of multiple-bit phase responses. For these multiple-bit RISs, the control degree of freedom remains one.

During the design process in the previous prototyping work of RIS-based 8PSK transmission\cite{RIS-based4}, the physical structure of the unit cell is carefully designed to achieve a large control range of phase response and a small fluctuation of amplitude response. That is, the design principle of the unit cell of this RIS aims to make (\ref{ss4}) insensitive to the change of $Z_{n,m}$, while enabling (\ref{ss5}) to have a large phase shift range. Consequently, we assume that the amplitude response of the RIS remains unchanged (e.g., ${A_{n,m}} = 1$) and the phase response can be flexibly regulated. Hence, the baseband of RIS-based MIMO can be converted into a constant envelope MIMO transmission model. The transmitted baseband signal in (\ref{ss13}) can be rewritten as
\begin{equation}\label{ss14}
{\mathbf{x}} = {\left[ {{e^{j{\varphi _{1,1}}}}, \dots ,{e^{j{\varphi _{1,M}}}}, \dots ,{e^{j{\varphi _{n,m}}}}, \dots ,{e^{j{\varphi _{N,M}}}}} \right]^T}.
\end{equation}
In the following, QAM modulation under the constant envelope constraint will be introduced and analyzed. The basic method is to use a non-linear modulation technique\cite{Independentcontrol}.
\vspace{-0.1cm}
\subsection{Basic Method}\label{BasicMethod}
Take the element ${{e^{j{\varphi _{n,m}}}}}$ in (\ref{ss14}) as an example, which represents the transmitted baseband symbol through the unit cell $U_{n,m}$. As discussed above, $U_{n,m}$ can only generate a modulation symbol with constant envelope at the carrier frequency, which results in the inability to achieve high-order modulation and limits the transmission rate. To unlock the constant envelope constraint, a non-linear modulation technique is proposed to realize high-order modulation which is described as follows. We define the baseband symbol as
\begin{equation}\label{ss15}
{s_{n,m}}(t){=}{e^{j{\varphi _{n,m}}(t)}}{=}\left\{ {\begin{array}{*{20}{cl}}
  {e^{j\frac{{\Delta \varphi }}{{{T_s}}}(t + {T_s} - {t_0})}}, & t \in \left[ {0,{t_0}} \right],\\
  {e^{j\frac{{\Delta \varphi }}{{{T_s}}}(t - {t_0})}}, & t \in \left( {{t_0},{T_s}} \right],
\end{array}} \right.
\end{equation}
in which the phase response ${\varphi _{n,m}}(t)$ changes linearly with time, $t_0$ is the circular time shift, $\frac{\Delta\varphi}{{{T_s}}}$ characterizes the changing rate of the phase that varies linearly with time, and ${T_s}$ is the symbol period. It is worth noting that the baseband symbol described in (\ref{ss15}) has two degrees of freedom: $t_0$ and $\Delta\varphi$, which enables QAM modulation at the harmonic frequencies.

{\textbf{Theorem 2.}} When the baseband symbol is defined as $s(t) = {e^{j\frac{{\Delta \varphi }}{{{T_s}}}(t + {T_s} - {t_0})}}$ for $t \in \left[ {0,{t_0}} \right]$, and $s(t) = {e^{j\frac{{\Delta \varphi }}{{{T_s}}}(t - {t_0})}}$ for $t \in \left( {{t_0},{T_s}} \right]$, its exponential Fourier series expansion gives
\begin{align}
s(t)=& \sum\limits_{l =  - \infty }^\infty  {{a_l}{e^{jl\frac{{2\pi }}{{{T_s}}}t}}}\notag\\
=& \sum\limits_{l =  - \infty }^\infty  \left| {\operatorname{sinc} \left(\frac{{\Delta \varphi }}{2} - l\pi \right)} \right|\notag\\
&\times {e^{j\left( { - l\frac{{2\pi {t_0}}}{{{T_s}}}{+}\frac{{\Delta \varphi }}{2}{-}l\pi{+}\bmod(\left\lfloor {\frac{{\Delta \varphi }}{{2\pi }}{-}l} \right\rfloor ,2) \cdot \pi{+}\varepsilon (2l\pi{-}\Delta \varphi) \cdot \pi } \right)}}\notag\\
&\times {e^{jl\frac{2\pi}{T_s}t}},\label{ss16}
\end{align}
where $\operatorname{sinc} (\cdot )$, $\bmod (\cdot )$, $\left\lfloor  \cdot  \right\rfloor$, and $\varepsilon ( \cdot )$ represent sinc function, modulus function, the round-down function, and the step function, respectively. In addition, $\left| {\operatorname{sinc} (\frac{{\Delta \varphi }}{2}{-}l\pi )} \right|$ is the amplitude of the $l^{th}$ order harmonic component and ${\left({- l\frac{{2\pi {t_0}}}{{{T_s}}}{+}\frac{{\Delta \varphi }}{2}{-}l\pi{+}\bmod(\left\lfloor {\frac{{\Delta \varphi }}{{2\pi }}{-}l}\right\rfloor ,2) \cdot \pi{+}\varepsilon (2l\pi{-}\Delta \varphi) \cdot \pi}\right)}$ represents the phase of the $l^{th}$ order harmonic component.

\emph{Proof}: See Appendix B. \hfill $\blacksquare$

Theorem 2 illustrates that the amplitude and phase of the harmonics can be adjusted independently, by changing the phase response of the unit cell of the RIS linearly with time during a symbol period. In particular, the amplitude modulation of the $l^{th}$ order harmonic can be achieved by adjusting $\Delta \varphi$. At the same time, the phase modulation of the $l^{th}$ order harmonic can be realized by adjusting $t_0$. Therefore, by manipulating the two degrees of freedom ($\Delta \varphi$ and $t_0$) in different symbol periods, QAM modulation can be realized on the harmonics.

For example, we can achieve QAM modulation on the $1^{st}$ order harmonic, where $f = {f_c} + \frac{1}{{{T_s}}}$. In this case, $l = 1$, and therefore, we have
\begin{equation}\label{ss17}
\left| {{a_1}} \right| = \left| {\operatorname{sinc} \left(\frac{{\Delta \varphi }}{2} - \pi \right)} \right|,
\end{equation}
and
\begin{equation}\label{ss18}
\begin{aligned}
\angle {a_1} =& -\frac{2\pi {t_0}}{T_s}{+}\frac{{\Delta \varphi }}{2}{-}\pi{+}\bmod \left(\left\lfloor {\frac{{\Delta \varphi }}{2\pi}{-}1} \right\rfloor, 2\right) \cdot \pi\\
&{+}\varepsilon (2\pi{-}\Delta \varphi) \cdot \pi .
\end{aligned}
\end{equation}

\begin{figure}
\centering
\includegraphics[scale = 0.625]{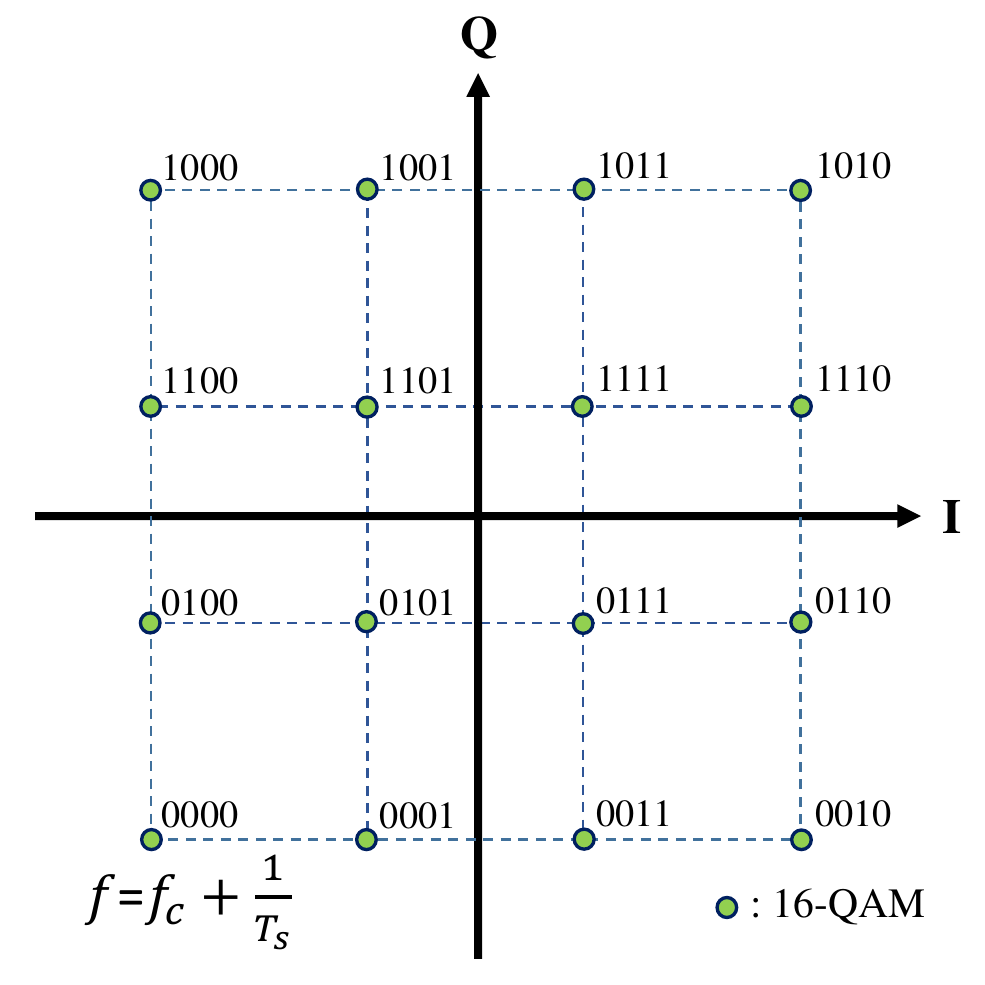}
\caption{An RIS-based 16-QAM modulation on the $1^{st}$ order harmonic with constant envelope constraint.}
\label{RISQAMmodulation}
\vspace{-0.2cm}
\end{figure}

As shown in Fig. \ref{RISQAMmodulation}, 16-QAM modulation can be performed on the $1^{st}$ order harmonic based on (\ref{ss17}) and (\ref{ss18}). Since the modulation is implemented on the harmonic, rather than the usual carrier signal, we refer to this modulation technique as a kind of \textbf{\emph{non-linear modulation}}. The corresponding mapping method of performing 16-QAM on the $1^{st}$ order harmonic is summarized in Table \ref{16QAMmappingsummary}. For instance, if the source bits `0010' need to be transmitted in a certain symbol duration, then we can set $t_0$ to 0.125$T_s$ and ${\Delta \varphi }$ to $2\pi$ to achieve the modulation.

\begin{table*}
\centering
\footnotesize
\caption{The mapping for performing 16-QAM on the $1^{st}$ order harmonic.}\label{16QAMmappingsummary}
\begin{tabular}{|c|c|c|c|c|c|}
\hline
\textbf{Symbol} & \textbf{Source Bits}  & \textbf{$\left| {{a_1}} \right|$} & \textbf{$\angle {a_1}$} & \textbf{${t_0}$} & \textbf{${\Delta \varphi }$}\\
\hline
0 &  0000 & 1 &  $\frac{5}{4}\pi$ & 0.375$T_s$ & $2\pi$\\
\hline
1 &  0001 & $\sqrt{\frac{5}{9}}$ & $\frac{3}{2}\pi-\arctan (\frac{1}{3})$ & 0.0962$T_s$ & ${\text{1.180}}\pi$\\
\hline
2 &  0010 & 1 & $\frac{7}{4}\pi$ & 0.125$T_s$ & $2\pi$\\
\hline
3 &  0011 & $\sqrt{\frac{5}{9}}$ &  $\frac{3}{2}\pi+\arctan (\frac{1}{3})$ & 0.994$T_s$ & ${\text{1.180}}\pi$\\
\hline
4 &  0100 & $\sqrt{\frac{5}{9}}$ & $\pi+\arctan (\frac{1}{3})$ & 0.244$T_s$ & ${\text{1.180}}\pi$\\
\hline
5 &  0101 & $\frac{1}{3}$ & $\frac{5}{4}\pi$ & 0.0123$T_s$ & ${\text{0.549}}\pi$\\
\hline
6 &  0110 & $\sqrt{\frac{5}{9}}$ & $2\pi-\arctan (\frac{1}{3})$ & 0.846$T_s$ & ${\text{1.180}}\pi$\\
\hline
7 &  0111 & $\frac{1}{3}$ & $\frac{7}{4}\pi$ & 0.762$T_s$ & ${\text{0.549}}\pi$\\
\hline
8 &  1000 & 1 & $\frac{3}{4}\pi$  & 0.625$T_s$ & $2\pi$\\
\hline
9 &  1001 & $\sqrt{\frac{5}{9}}$ & $\frac{1}{2}\pi+\arctan (\frac{1}{3})$ & 0.494$T_s$ & ${\text{1.180}}\pi$\\
\hline
10 &  1010 & 1 & $\frac{1}{4}\pi$ & 0.875$T_s$ & $2\pi$\\
\hline
11 &  1011 & $\sqrt{\frac{5}{9}}$ & $\frac{1}{2}\pi-\arctan (\frac{1}{3})$ & 0.596$T_s$ & ${\text{1.180}}\pi$\\
\hline
12 &  1100 & $\sqrt{\frac{5}{9}}$ & $\pi-\arctan (\frac{1}{3})$ & 0.346$T_s$ & ${\text{1.180}}\pi$\\
\hline
13 &  1101 & $\frac{1}{3}$ &  $\frac{3}{4}\pi$ & 0.262$T_s$ & ${\text{0.549}}\pi$\\
\hline
14 &  1110 & $\sqrt{\frac{5}{9}}$ & $\arctan (\frac{1}{3})$ & 0.744$T_s$ & ${\text{1.180}}\pi$\\
\hline
15 &  1111 & $\frac{1}{3}$ & $\frac{1}{4}\pi$ & 0.512$T_s$ & ${\text{0.549}}\pi$\\
\hline
\end{tabular}
\end{table*}
Theorem 2 gives a method for realizing QAM modulation under the constant envelope constraint, which enables RIS-based MIMO-QAM transmission. By replacing ${\mathbf{x}}$ with ${\mathbf{s}}$ in (\ref{ss13}), the baseband expression of RIS-based MIMO-QAM transmission can be obtained as
\begin{equation}\label{ss19}
{\mathbf{y}} = \sqrt p {\mathbf{Hs}} + {\mathbf{n}},
\end{equation}
where ${\mathbf{s}}= {\left[ {{s_{1,1}},\dots ,{s_{1,M}}, \dots ,{s_{n,m}}, \dots ,{s_{N,M}}} \right]^T}\in {\mathbb{C}^{NM \times 1}}$, with $s_{n,m}$ being the transmitted symbol through the unit cell $U_{n,m}$, which has been defined in (\ref{ss15}).
\vspace{-0.1cm}
\subsection{Beamforming and Gain}\label{BeamformingGain}
In the above subsection, the basic method of RIS-based MIMO-QAM transmission under the constant envelope constraint is introduced. Under the proposed non-linear modulation technique, the RIS-based transmitter can also achieve the beamforming functionality. In particular, the transmitted baseband signal described in (\ref{ss14}) can be further expressed as
\begin{equation}\label{ss20}
\begin{aligned}
{\mathbf{x}} =& \left[ {e^{j\varphi _{_{1,1}}^{beam}}}{s_{1,1}},{e^{j\varphi _{_{1,2}}^{beam}}}{s_{1,2}}, \dots ,{e^{j\varphi _{_{1,M}}^{beam}}}{s_{1,M}},\right.\\
&\left.\dots ,{e^{j\varphi _{_{n,m}}^{beam}}}{s_{n,m}}, \dots ,{e^{j\varphi _{_{N,M}}^{beam}}}{s_{N,M}} \right]^T\\
=& \mathbf{\Phi _{beam}}{\left[ {{s_{1,1}},{s_{1,2}}, \dots ,{s_{1,M}}, \dots ,{s_{n,m}}, \dots ,{s_{N,M}}} \right]^T}\\
=&\mathbf{\Phi _{beam}}{\mathbf{s}},
\end{aligned}
\end{equation}
where
\begin{equation}\label{ss21}
\begin{aligned}
{\mathbf{\Phi}}_{\mathbf{beam}}{=}{\rm diag} &\left\{{e^{j\varphi _{_{1,1}}^{beam}}},{e^{j\varphi _{_{1,2}}^{beam}}}, \dots ,{e^{j\varphi _{_{1,M}}^{beam}}},\right.\\
&\left.\dots ,{e^{j\varphi _{_{n,m}}^{beam}}}, \dots ,{e^{j\varphi _{_{N,M}}^{beam}}}\right\}\in{\mathbb{C}^{NM \times NM}},
\end{aligned}
\end{equation}
is the beamforming matrix, in which ${\varphi _{_{n,m}}^{beam}}$ is the beamforming factor of the unit cell $U_{n,m}$. By substituting (\ref{ss20}) into (\ref{ss13}), the baseband expression of RIS-based MIMO wireless communication system can be rewritten as
\begin{equation}\label{ss22}
{\mathbf{y}} = \sqrt p {\mathbf{H}}{{\mathbf{\Phi }}_{{\mathbf{beam}}}}{\mathbf{s}} + {\mathbf{n}},
\end{equation}
which gives the general baseband expression of RIS-based MIMO, for high-order modulation and beamforming simultaneously with constant envelope constraint.

For analytical convenience, in the sequel, we consider free-space propagation, and assume that the receiving antennas are in the far-field of the RIS and the transmitting signal is beamformed to the $k^{th}$ receiving antenna. According to (\ref{ss8}) and (\ref{ss9}), when ${{\mathbf{\Phi }}_{{\mathbf{beam}}}}$ is designed to align the received signals of the $k^{th}$ receiving antenna from all the unit cells, the received signal of the $k^{th}$ receiving antenna can be expressed as (\ref{ss23}) shown at the top of the next page,
\begin{figure*}[!t]
\begin{align}\label{ss23}
{y_k}&=\sum\limits_{m = 1}^M {\sum\limits_{n = 1}^N {\frac{{\sqrt {G{G_r}{\lambda ^2}F\left( {\theta _{n,m}^{AOD,k},\phi _{n,m}^{AOD,k}} \right){F^{rx}}\left( {\theta _{n,m}^{AOA,k},\phi _{n,m}^{AOA,k}} \right)S{d_x}{d_y}} }}{{4\pi d_{n,m}^k}}{e^{\frac{{ - j2\pi d_{n,m}^k}}{\lambda }}}{e^{j\varphi _{_{n,m}}^{beam}}}{s_{n,m}}{e^{j2\pi {f_c}t}}} }\notag\\
  &\approx \sum\limits_{m = 1}^M {\sum\limits_{n = 1}^N {\frac{{\sqrt {G{G_r}{\lambda ^2}F\left( {{\theta ^{AOD,k}},{\phi ^{AOD,k}}} \right){F^{rx}}\left( {{\theta ^{AOA,k}},{\phi ^{AOA,k}}} \right)S{d_x}{d_y}} }}{{4\pi {d^k}}}{e^{\frac{{ - j2\pi d_{n,m}^k}}{\lambda }}}{e^{j\varphi _{_{n,m}}^{beam}}}{s_{n,m}}{e^{j2\pi {f_c}t}}} }\notag \\
  &= \frac{{NM\sqrt {G{G_r}{\lambda ^2}F\left( {{\theta ^{AOD,k}},{\phi ^{AOD,k}}} \right){F^{rx}}\left( {{\theta ^{AOA,k}},{\phi ^{AOA,k}}} \right)S{d_x}{d_y}} }}{{4\pi {d^k}}}{s}{e^{j2\pi {f_c}t}}
\end{align}
\hrulefill
\end{figure*}
\begin{figure*}
\centering
\resizebox{6in}{!}
{
\begin{tabular}{ccc}
\includegraphics*{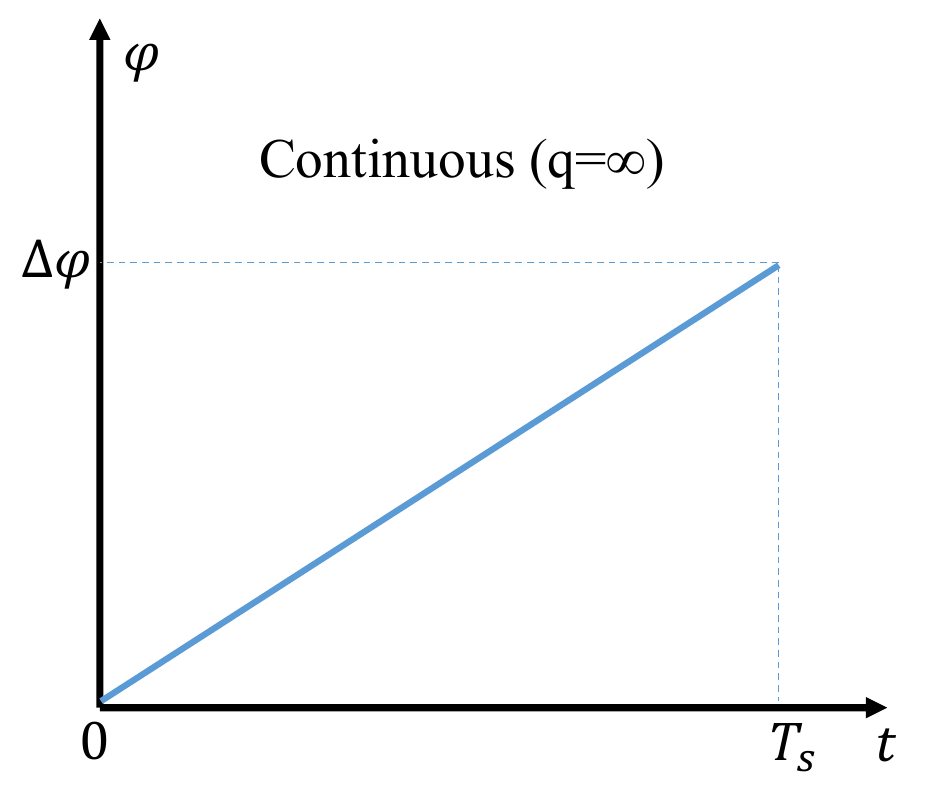} & \includegraphics*{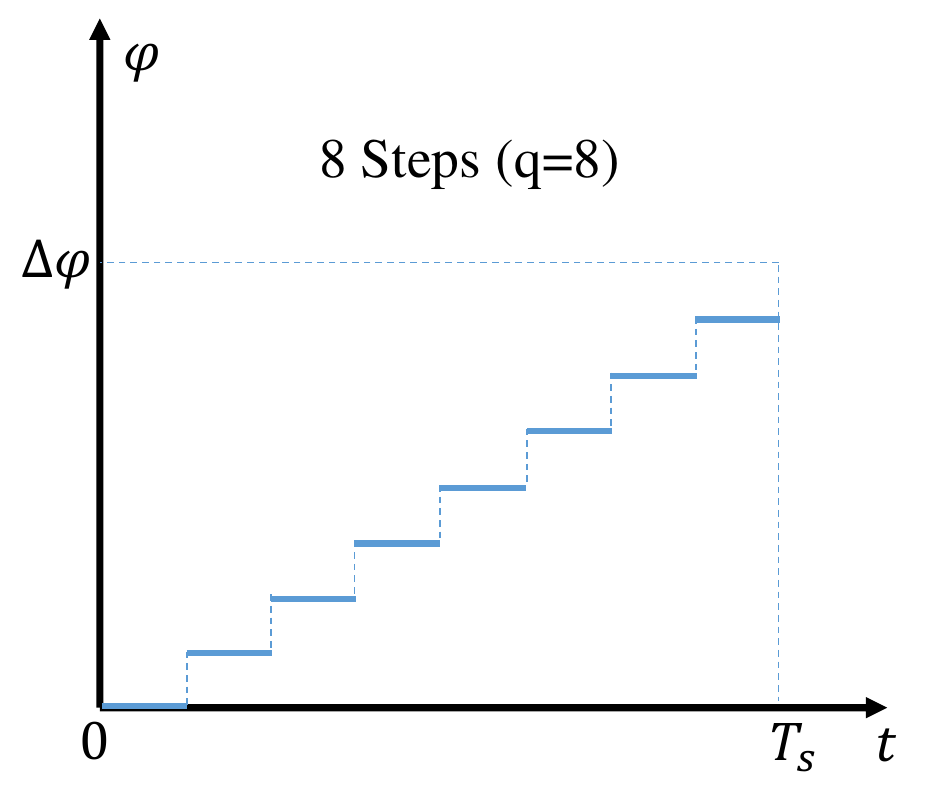} & \includegraphics*{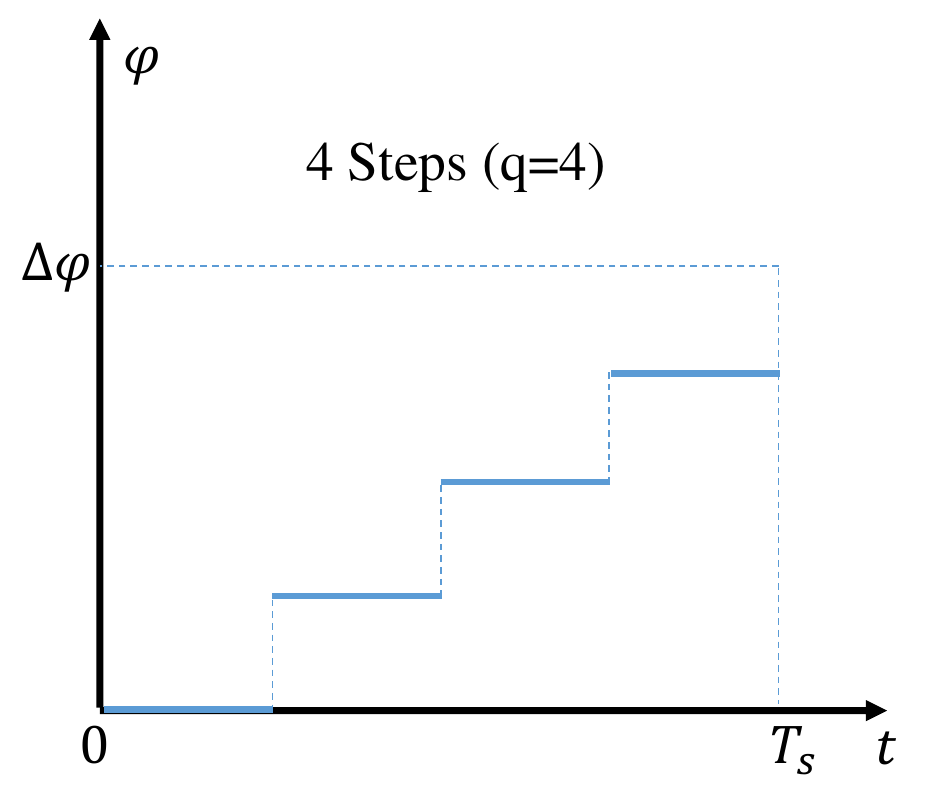} \\
\Large (a) & \Large (b) & \Large (c)\\
\end{tabular}
}
\caption{Different discrete phase shift steps in the proposed RIS-based QAM symbol. (a) $q$=$\infty$. (b) $q$=8. (c) $q$=4.}
\label{RISDiscretePhase}
\end{figure*}
where $d^k$, $\theta^{AOD,k}$, $\phi^{AOD,k}$, $\theta^{AOA,k}$ and $\phi^{AOA,k}$ represent the distance between the center of the RIS and the $k^{th}$ receiving antenna, the elevation angle and the azimuth angle from the center of the RIS to the $k^{th}$ receiving antenna, and the elevation angle and the azimuth angle from the $k^{th}$ receiving antenna to the center of the RIS, respectively. The beamforming factor $ {e^{j\varphi _{_{n,m}}^{beam}}}$ is equal to ${e^{\frac{{j2\pi d_{n,m}^k}}{\lambda }}}$, i.e., ${{\mathbf{\Phi }}_{{\mathbf{beam}}}} = {\rm{diag}}\left\{{e^{\frac{{j2\pi d_{1,1}^k}}{\lambda }}}, \dots ,{e^{\frac{{j2\pi d_{1,M}^k}}{\lambda }}}, \dots ,{e^{\frac{{j2\pi d_{n,m}^k}}{\lambda }}}, \dots ,{e^{\frac{{j2\pi d_{N,M}^k}}{\lambda }}}\right\}$. Meanwhile, all the unit cells transmit the same symbol $s$.

Equation (\ref{ss23}) reveals that the beamforming gain of RIS-based transmitter is proportional to the number of the unit cells of the RIS (i.e., $NM$), as well as the square root of the size of the unit cell (i.e., $\sqrt {{d_x}{d_y}}$). In other words, the larger the aperture of RIS, the higher the beamforming gain, which is consistent with the intuition.

\section{System Design and Analysis}\label{SystemDesignAnalysis}
In the rest of this paper, we design and implement an RIS-based 2$\times$2 MIMO-QAM wireless communication system to validate the proposed method in Section III. In this section, we first discuss the impacts of the hardware constraints of the RIS on the system design, and then give the detailed design of the RIS-based transmitter and the receiver in the RIS-based 2$\times$2 MIMO-QAM wireless communication prototype. The technique is generalizable for any size of MIMO.
\vspace{-0.1cm}
\subsection{Analyses of Hardware Constraints}\label{ImperfectFactors}

\subsubsection{Phase Dependent Amplitude}\label{PhaseDependentAmplitude}
The method we propose to achieve RIS-based MIMO-QAM transmission assumes that the EM response of the unit cells of the RIS behaves in a constant envelope manner, i.e., the phase response can be flexibly controlled, and the amplitude response is constant. However, the practical amplitude response of the unit cells is not strictly constant since the reflection amplitude and phase responses of the unit cell are usually strongly coupled. During the design process of the unit cell structure, researchers tend to make the fluctuation of the amplitude response as small as possible, while obtaining a sufficiently large range for phase response control \cite{Amp1,Amp2,Amp3}. It is worth noting that our proposed method of achieving RIS-based QAM is robust under the phase dependent amplitude hardware constraint of the RIS.

Take the phase dependent amplitude response described below as an example, i.e.,
\begin{equation}\label{ss24}
A\left( \varphi  \right) = \left\{ {\begin{array}{*{20}{cl}}
  0.7 + \frac{{0.3}}{\pi }\,\varphi , & \varphi  \in \left[ {0,\pi } \right), \\
  1.3 - \frac{{0.3}}{\pi }\,\varphi , &\varphi  \in \left[ {\pi ,2\pi } \right],
\end{array}} \right.
\end{equation}
whose maximal amplitude fluctuation is 3 dB ($20\log 0.7 = - 3$ dB). Considering the baseband symbol defined in Theorem 2 with $t_0=0$, we have
\begin{equation}\label{ss25}
\begin{aligned}
s(t) =& A\left( {\frac{{\Delta \varphi }}{{{T_s}}}t} \right){e^{j\left( {\frac{{\Delta \varphi }}{{{T_s}}}t} \right)}} = \sum\limits_{l =  - \infty }^\infty  {{a_l}{e^{jl\frac{{2\pi }}{{{T_s}}}t}}}\\
=& \sum\limits_{l =  - \infty }^\infty  \left(\frac{1}{T_s}\int_0^{T_s} A\left(\frac{\Delta \varphi}{T_s}\tau\right)e^{j\left(\frac{\Delta \varphi}{T_s}\tau \right)}e^{- jl\frac{2\pi}{T_s}\tau}d\tau\right)\\
&\times e^{jl\frac{2\pi}{T_s}t}, ~t \in \left[ {0,{T_s}} \right].
\end{aligned}
\end{equation}
Letting $l=1$, the $1^{st}$ order harmonic component of $s(t)$ is given by
\begin{equation}\label{ss26}
\begin{aligned}
{a_1} &= \frac{1}{{{T_s}}}\int_0^{{T_s}} {A\left( {\frac{{\Delta \varphi }}{{{T_s}}}t} \right){e^{j\left( {\frac{{\Delta \varphi }}{{{T_s}}}t} \right)}}{e^{ - j\frac{{2\pi }}{{{T_s}}}t}}dt}\\
&= \frac{{1}}{{{T_s}}}\int_0^{{T_s}} {A\left( {\frac{{\Delta \varphi }}{{{T_s}}}t} \right){e^{j\left( {\frac{{\Delta \varphi }}{{{T_s}}} - \frac{{2\pi }}{{{T_s}}}} \right)t}}dt}.
\end{aligned}
\end{equation}

As the amplitude response $A\left( \varphi  \right) > 0$, (\ref{ss26}) is maximized when $\Delta \varphi  = 2\pi$. Then we have
\begin{equation}\label{ss27}
\begin{aligned}
&a_1^{\max }\\
&= \frac{{1}}{{{T_s}}}\int_0^{{T_s}} {A\left( {\frac{{2\pi }}{{{T_s}}}t} \right)dt}\\
&= \frac{{1}}{{{T_s}}}\left[ {\int_0^{\frac{{{T_s}}}{2}} {\left(0.7{+}\frac{{0.6}}{{{T_s}}}t\right)dt}{+} \int_{\frac{{{T_s}}}{2}}^{{T_s}} {\left(1.3{-}\frac{{0.6}}{{{T_s}}}t\right)dt} } \right]\\
&= 0.85,
\end{aligned}
\end{equation}
based on which we can realize the symbols `0', `2', `8', and `10' of 16-QAM by setting the circular time shift $t_0$ to $0.375T_s$, $0.125T_s$, $0.625T_s$, and $0.875T_s$, respectively, according to the time delay property of the Fourier transform. For other symbols of 16-QAM, take symbol `5' as an example here. Its corresponding value of $\Delta \varphi$ can be obtained by solving the equation $\left| {{a_1}} \right| = \frac{a_1^{\max }}{3}=\frac{{0.85}}{3}$, and then use the time delay property of the Fourier transform to design $t_0$ such that $\angle {a_1} = \frac{5}{4}\pi$. For the various phase dependent amplitude functions in practice, QAM can be achieved through the above design method and process.

\subsubsection{Discrete Phase Shift}\label{DiscretePhaseShift}

\begin{figure}
\centering
\includegraphics[scale = 0.54]{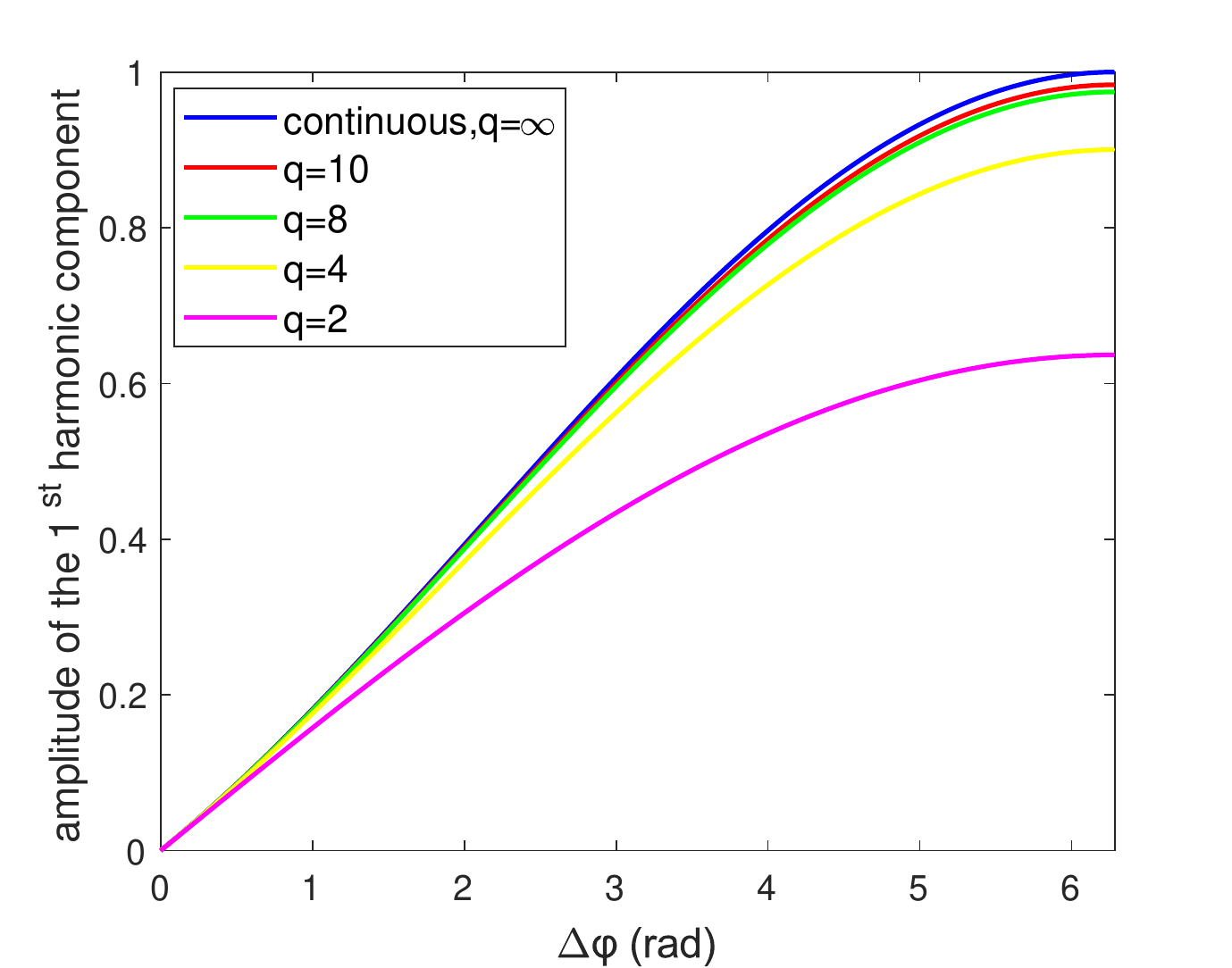}
\caption{The amplitude of the $1^{st}$ order harmonic component with different discrete phase shift steps.}
\label{AmplitudeDifferent}
\vspace{-0.2cm}
\end{figure}
The baseband symbol we design in Theorem 2 to realize RIS-based QAM modulation assumes that the phase response of the unit cells can vary continuously with time as shown in Fig. \ref{RISDiscretePhase}(a). However, such an ideal signal does not exist in practice when facing system implementation. This is because the phase response of the unit cell is controlled by an external control signal, which is generated by the DAC with discrete output characteristic. Let $q$ denote the number of the discrete phase shift steps in a practical baseband symbol of RIS-based QAM. For example, $q$ is equal to $\infty$, 8, and 4 in Fig. \ref{RISDiscretePhase}(a), (b), and (c), respectively. It is worth noting that although the baseband symbol of RIS-based QAM as ideal as Fig. \ref{RISDiscretePhase}(a) can be achieved when the phase response of the unit cell is controlled by a high-resolution DAC, the symbol rate of RIS-based QAM will be significantly limited. The maximal symbol rate of our proposed RIS-based QAM is limited by
\begin{equation}\label{ss28}
R_{symbol}^{\max } = \frac{{{R_{DAC}}}}{q},
\end{equation}
where $R_{DAC}$ is the maximum sampling rate of the DAC. For instance, when $R_{DAC}=100$ MSa/s and $q=1000$, the maximal symbol rate is limited as $100$kS/s, which is relatively a low symbol rate. In contrast, a high symbol rate can be obtained when the value of $q$ is small. The impact of discrete phase shift on the baseband symbol of RIS-based QAM is analyzed next.

When considering discrete phase shift and letting $t_0=0$, the baseband symbol of RIS-based QAM in Theorem 2 is redefined as
\begin{equation}\label{ss29}
\widetilde s(t) = {e^{j\frac{{\vartriangle \varphi }}{q}p}},~\mbox{for }t \in \left[ {\frac{{Ts}}{q}p,\frac{{Ts}}{q}(p + 1)} \right),
\end{equation}
where $p \in [0,1, \dots ,q - 1]$. The $l^{th}$ order harmonic component of $\widetilde s(t)$ is
\begin{equation}\label{ss30}
\begin{aligned}
&\widetilde {{a_l}}= \frac{1}{{{T_s}}}\int_0^{{T_s}} {\widetilde s(t){e^{ - jl\frac{{2\pi }}{{{T_s}}}t}}dt}\\
&= \sum\limits_{p = 0}^{q - 1} {\frac{1}{{{T_s}}}\int_{\frac{{pTs}}{q}}^{\frac{{(p + 1)Ts}}{q}} {{e^{j\frac{{\Delta \varphi }}{q}p}}{e^{ - jl\frac{{2\pi }}{{{T_s}}}t}}dt} }\\
&= \frac{{j\left({e^{ - jl\frac{{2\pi }}{q}}}{-}1\right)}}{{l2\pi }}\frac{{1{-}{e^{j(\Delta \varphi  - l2\pi )}}}}{{1{-}{e^{j\left(\frac{{\Delta \varphi  - l2\pi }}{q}\right)}}}}\\
&= \frac{{2\sin \left(\frac{{l\pi }}{q}\right){e^{ - j\frac{{l\pi }}{q}}}\sin \left(\frac{{\Delta \varphi }}{2}{-} l\pi \right){e^{j\left(\frac{{\Delta \varphi }}{2}{-}l\pi \right)}}}}{{l2\pi \sin \left(\frac{{\frac{{\Delta \varphi }}{2}{-}l\pi }}{q}\right){e^{j\left(\frac{{\frac{{\Delta \varphi }}{2}{-}l\pi }}{q}\right)}}}}\\
&= \frac{{\operatorname{sinc} \left(\frac{{l\pi }}{q}\right)}}{{\operatorname{sinc} \left(\left( {\frac{{\Delta \varphi }}{2}{-}l\pi } \right)\frac{1}{q}\right)}}\operatorname{sinc} \left(\frac{{\Delta \varphi }}{2}{-}l\pi \right){e^{j\left( {\frac{{\Delta \varphi }}{2}{-}l\pi{-}\frac{{\Delta \varphi }}{{2q}}} \right)}}.
\end{aligned}
\end{equation}


Meanwhile, according to Theorem 2 and (\ref{sss15}), the $l^{th}$ order harmonic component of the ideal baseband symbol $s(t)$ without discrete phase shift is
\begin{equation}\label{ss31}
\begin{aligned}
{a_l}{=}\frac{1}{{{T_s}}}\int_0^{{T_s}}{{e^{j(\frac{{\Delta \varphi }}{{{T_s}}}{-}l\frac{{2\pi }}{{{T_s}}})t}}dt}  {=}\operatorname{sinc} \left(\frac{{\Delta \varphi }}{2}{-}l\pi \right){e^{j(\frac{{\Delta \varphi }}{2}{-}l\pi )}}.
\end{aligned}
\end{equation}

By comparing (\ref{ss30}) and (\ref{ss31}), we have
\begin{equation}\label{ss32}
\frac{{{{\widetilde a}_l}}}{{{a_l}}} = \frac{{\operatorname{sinc} (\frac{{l\pi }}{q})}}{{\operatorname{sinc} \left(\left( {\frac{{\Delta \varphi }}{2} - l\pi } \right)\frac{1}{q}\right)}}{e^{ - j\frac{{\Delta \varphi }}{{2q}}}},
\end{equation}
from which we can see that the larger the number of the discrete phase shift steps $q$, the smaller the impact of the discrete phase shift on the baseband symbol of RIS-based QAM. In particular, Fig. \ref{AmplitudeDifferent} shows the amplitude of the $1^{st}$ order harmonic component with different discrete phase shift steps. As $q$ increases, $\left| {{{\widetilde a}_1}} \right|$ with discrete phase shift steps quickly approaches the ideal $\left| {{{a}_1}} \right|$ without discretization. As shown in Fig. \ref{AmplitudeDifferent}, the impact of the discrete phase shift is already small enough ($\left| {{{\widetilde a}_1}} \right|=0.9745$) when $q= 8$. Therefore, our proposed method of achieving RIS-based QAM is robust with the discrete phase shift steps, and a high symbol rate can be obtained.

\vspace{-0.1cm}
\subsection{Transmitter Design}\label{Transmitter}
The RIS we implemented has 256 unit cells ($N=32$ and $M=8$). In principle, if each unit cell is controlled by a dedicated DAC, the RIS-based transmitter can achieve simultaneous transmission of different signals over 256 unit cells. The use of only two DACs here to control the unit cells of the RIS is mainly limited by our experimental hardware conditions. Therefore, an RIS-based 2$\times$2 MIMO-QAM wireless communication system is designed and implemented. Nevertheless, it should be sufficient to demonstrate the great potential of realizing UM-MIMO and holographic MIMO technologies through the RISs.

\begin{figure*}
\centering
\includegraphics[scale = 0.65]{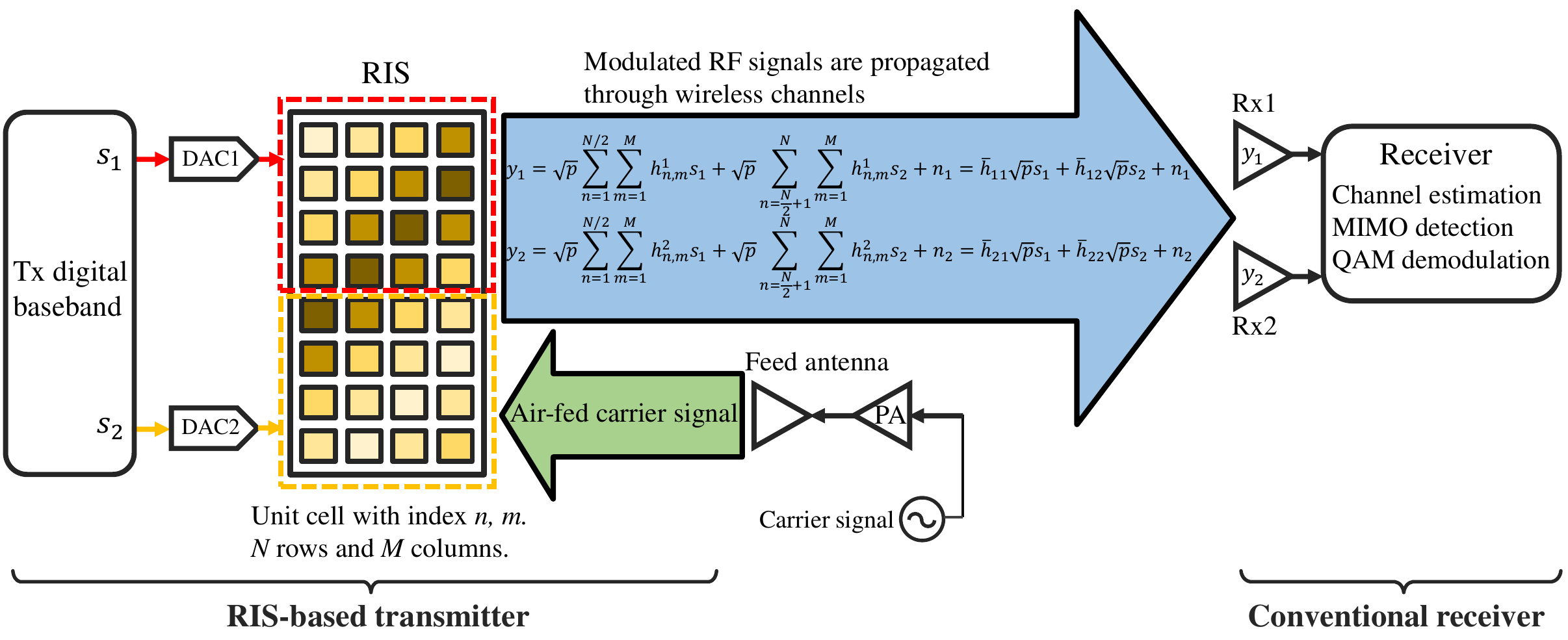}
\caption{The diagram of RIS-based 2$\times$2 MIMO-QAM wireless communication system.}
\label{RISDiagramSystem}
\vspace{-0.2cm}
\end{figure*}

\begin{figure}
\centering
\includegraphics[scale = 0.52]{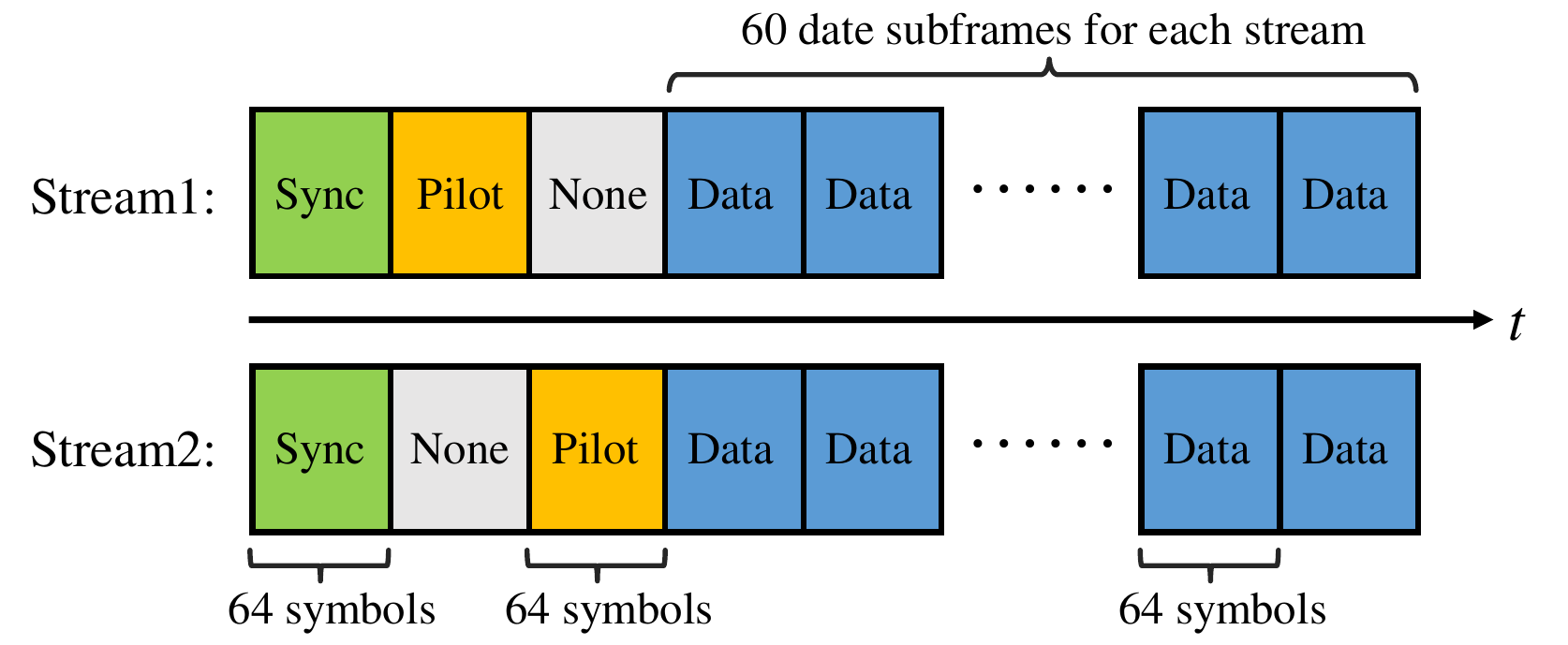}
\caption{The wireless frame structure of RIS-based 2$\times$2 MIMO-QAM wireless communication system.}
\label{RISWirelseeFrame}
\vspace{-0.2cm}
\end{figure}

The diagram of the RIS-based 2$\times$2 MIMO-QAM wireless communication system is presented in Fig. \ref{RISDiagramSystem}. One bit stream is transmitted by half of the RIS (the red half shown in Fig. \ref{RISDiagramSystem}) and the other bit stream is transmitted by the other half (the orange half shown in Fig. \ref{RISDiagramSystem}). According to the baseband expression of RIS-based MIMO-QAM transmission expressed by (\ref{ss19}), we have
\begin{equation}\label{ss33}
\begin{array}{*{20}{l}}
  {y_1} &{=} \sqrt p \sum\limits_{m = 1}^M \sum\limits_{n = 1}^{\frac{N}{2}} {h_{n,m}^1{s_1}}  + \sqrt p \sum\limits_{m = 1}^M \sum\limits_{n = \frac{N}{2} + 1}^N {h_{n,m}^1{s_2}}  + {n_1}\\
  &= \overline h {}_{11}{\sqrt p}{s_1} + \overline h {}_{12}{\sqrt p}{s_2} + {n_1}, \\
  {y_2} &{=} \sqrt p \sum\limits_{m = 1}^M \sum\limits_{n = 1}^{\frac{N}{2}} {h_{n,m}^2{s_1}}  + \sqrt p \sum\limits_{m = 1}^M \sum\limits_{n = \frac{N}{2} + 1}^N {h_{n,m}^2{s_2}}  + {n_2}\\
  &= \overline h {}_{21}{\sqrt p}{s_1} + \overline h {}_{22}{\sqrt p}{s_2} + {n_2},
\end{array}
\end{equation}
where $\overline h {}_{11}{=}{\sum\limits_{m = 1}^M {\sum\limits_{n = 1}^{\frac{N}{2}} {h_{n,m}^1} } }$, $\overline h {}_{12}{=}{\sum\limits_{m = 1}^M {\sum\limits_{n = \frac{N}{2} + 1}^N {h_{n,m}^1} } }$, $\overline h {}_{21}{=}{\sum\limits_{m = 1}^M {\sum\limits_{n = 1}^{\frac{N}{2}} {h_{n,m}^2} } }$, and $\overline h {}_{22}{=}{\sum\limits_{m = 1}^M {\sum\limits_{n = \frac{N}{2} + 1}^N {h_{n,m}^2} }}$. $\overline h _{11}$, $\overline h _{21}$, $\overline h _{12}$ and $\overline h _{22}$ represent the channel between the red half of the RIS and the first receiving antenna Rx1, the channel between the red half of the RIS and the second receiving antenna Rx2, the channel between the orange half of the RIS and the antenna Rx1, the channel between the orange half of the RIS and antenna Rx2, respectively. As can be seen from (\ref{ss33}), the signal expression of RIS-based 2x2 MIMO-QAM transmission is the same with that of the conventional one. Therefore, we designed a common wireless frame structure, which includes one synchronization subframe, one pilot subframe, and sixty data subframes as shown in Fig. \ref{RISWirelseeFrame}. The pilot subframe consists of 64 RIS-based BPSK symbols and each data subframe consists of 64 RIS-based 16-QAM symbols, which perform QAM modulation on the $1^{st}$ order harmonic. Every frame can transmit 30720 bits information ($2\times60\times64\times4=30720$). The pilot subframes of the two streams are orthogonal to each other in the time domain, so that $\overline h _{11}$, $\overline h _{21}$, $\overline h _{12}$ and $\overline h _{22}$ can be easily obtained at the receiver side.

\vspace{-0.1cm}
\subsection{Receiver Design}\label{Receiver}
As shown in Fig. \ref{RISDiagramSystem}, a conventional two-channel receiver is designed to recover the transmitted bitstreams. Since the received symbols are the RIS-based QAM symbols designed in Section III, we oversample each symbol by 8 times, i.e., 8 samples are sampled for each symbol. Then the fast Fourier transform (FFT) is carried out on the 8 samples of each RIS-based QAM symbol to calculate the amplitude and phase of the $1^{st}$ order harmonic, thus getting the raw QAM symbols. After synchronization, the receiver performs 2$\times$2 MIMO channel estimation, 2$\times$2 MIMO detection, and QAM demodulation. In particular, least square (LS) algorithm is used for channel estimation and zero forcing (ZF) is used for channel equalization. We use LS and ZF in the system design here because of their ease of implementation, which enables the quick validation of our proposed method for RIS-based MIMO-QAM transmission. Through the above demodulation process, the two bitstreams can be recovered.
\vspace{-0.1cm}
\subsection{{BER Performance}}\label{BER}
According to the baseband expression of the RIS-based MIMO-QAM transmission described in (\ref{ss19}), the proposed RIS-based MIMO system only differs in terms of the transmitter hardware architecture, while sharing the same essential principle and basic mathematical expression with the conventional MIMO system. As a result, the proposed RIS-based system has the same bit error rate (BER) performance with the conventional system under the same signal-to-noise ratio (SNR) condition. In particular, we take the RIS-based 2$\times$2 MIMO-QAM wireless communication system designed above as an example, and derive its theoretical BER in the following. In Section~V, we will compare the derived theoretical BER with the measured BER, thus experimentally verifying that the RIS-based transmission shares the same BER-SNR performance with the conventional system.

Assume that the 2$\times$2 MIMO wireless channel in (\ref{ss33}) remains constant during the transmission, i.e., $\overline h {}_{11}$, $\overline h {}_{12}$, $\overline h {}_{21}$ and $\overline h {}_{22}$ are considered to be complex constants.\footnote{Our prototype system is built in a simple indoor environment as shown in Fig. \ref{RISPhoto}, and the people, RIS-based transmitter and conventional receiver remain stationary during the measurement. As a result, the wireless channel is almost changeless during the experiment. Therefore, we assume that the 2$\times$2 MIMO wireless channel remains constant here for BER analysis, which enables the corresponding comparison with the measured BER in the next section. In the experiment, we average the 64 results of LS channel estimation of each frame to obtain the almost actual channel parameters ($\overline h {}_{11}$, $\overline h {}_{12}$, $\overline h {}_{21}$ and $\overline h {}_{22}$).} As described in the subsection of receiver design, FFT is performed on both sides of (\ref{ss33}) to obtain the raw QAM symbols on the $1^{st}$ order harmonic. According to the linearity property of FFT, we have
\begin{equation}\label{ss34}
\begin{array}{*{20}{c}}
  {{Y_1} = {{\overline h}_{11}}\sqrt p{S_1} + {{\overline h}_{12}}\sqrt p{S_2} + {W_1}}, \\
  {{Y_2} = {{\overline h}_{21}}\sqrt p{S_1} + {{\overline h}_{22}}\sqrt p{S_2} + {W_2}},
\end{array}
\end{equation}
where $Y_1$ and $Y_2$ are the received raw symbols on the $1^{st}$ order harmonic, $S_1$ and $S_2$ are the transmitted original symbols on the $1^{st}$ order harmonic satisfying $\mathbb{E}\left\{ {{{\left| {{S_1}} \right|}^2}} \right\} = \mathbb{E}\left\{ {{{\left| {{S_2}} \right|}^2}} \right\} = 1$, $\mathbb{E}\left\{  \cdot  \right\}$ represents taking expectation, and $W_1$ and $W_2$ are two independent complex Gaussian noises with zero mean and the same variance ${\sigma}^2 $ on the $1^{st}$ order harmonic of the two receiving chains, respectively. By applying ZF equalization algorithm on (\ref{ss34}), we have
\begin{equation}\label{ss35}
\begin{array}{*{20}{c}}
  {{\widetilde S_1} = {S_1} + \frac{{{{\bar h}_{22}}}}{{\sqrt p \left( {{{\bar h}_{11}}{{\bar h}_{22}} - {{\bar h}_{12}}{{\bar h}_{21}}} \right)}}{W_1} + \frac{{ - {{\bar h}_{12}}}}{{\sqrt p \left( {{{\bar h}_{11}}{{\bar h}_{22}} - {{\bar h}_{12}}{{\bar h}_{21}}} \right)}}{W_2}}, \\
  {{\widetilde S_2} = {S_2} + \frac{{ - {{\bar h}_{21}}}}{{\sqrt p \left( {{{\bar h}_{11}}{{\bar h}_{22}} - {{\bar h}_{12}}{{\bar h}_{21}}} \right)}}{W_1} + \frac{{{{\bar h}_{11}}}}{{\sqrt p \left( {{{\bar h}_{11}}{{\bar h}_{22}} - {{\bar h}_{12}}{{\bar h}_{21}}} \right)}}{W_2}},
\end{array}
\end{equation}
where $\widetilde S_1$ and $\widetilde S_2$ are the recovered symbols of stream1 and stream2, respectively.

According to (\ref{ss35}), the signal-to-noise ratios of stream1 and stream2 under ZF equalization algorithm are respectively
\begin{equation}\label{ss36}
\begin{array}{*{20}{l}}
\text {SNR}_1^{\text{ZF}}&= \frac{{\mathbb{E}\left\{ {{{\left| {{S_1}} \right|}^2}} \right\}}}{{\left({{\left| {\frac{{{{\bar h}_{22}}}}{{\sqrt p \left( {{{\bar h}_{11}}{{\bar h}_{22}} - {{\bar h}_{12}}{{\bar h}_{21}}} \right)}}} \right|}^2} + {{\left| {\frac{{ - {{\bar h}_{12}}}}{{\sqrt p \left( {{{\bar h}_{11}}{{\bar h}_{22}} - {{\bar h}_{12}}{{\bar h}_{21}}} \right)}}} \right|}^2}\right){\sigma ^2}}}\\
&= \frac{{p{{\left| {{{\bar h}_{11}}{{\bar h}_{22}} - {{\bar h}_{12}}{{\bar h}_{21}}} \right|}^2}}}{{\left( {{{\left| {{{\bar h}_{22}}} \right|}^2} + {{\left| {{{\bar h}_{12}}} \right|}^2}} \right){\sigma ^2}}}, \\
\text {SNR}_2^{\text{ZF}}&= \frac{{\mathbb{E}\left\{ {{{\left| {{S_2}} \right|}^2}} \right\}}}{{\left({{\left| {\frac{{ - {{\bar h}_{21}}}}{{\sqrt p \left( {{{\bar h}_{11}}{{\bar h}_{22}} - {{\bar h}_{12}}{{\bar h}_{21}}} \right)}}} \right|}^2} + {{\left| {\frac{{{{\bar h}_{11}}}}{{\sqrt p \left( {{{\bar h}_{11}}{{\bar h}_{22}} - {{\bar h}_{12}}{{\bar h}_{21}}} \right)}}} \right|}^2}\right){\sigma ^2}}}\\
&= \frac{{p{{\left| {{{\bar h}_{11}}{{\bar h}_{22}} - {{\bar h}_{12}}{{\bar h}_{21}}} \right|}^2}}}{{\left( {{{\left| {{{\bar h}_{21}}} \right|}^2} + {{\left| {{{\bar h}_{11}}} \right|}^2}} \right){\sigma ^2}}}.
\end{array}
\end{equation}
Since $S_1$ and $S_2$ use the standard 16-QAM mapping and Gray coding scheme as shown in Fig. \ref{RISQAMmodulation}, the utilized QAM demodulation method here is based on Euclidean distance, which is widely used in conventional systems. In addition, (\ref{ss35}) reveals that the QAM demodulation is performed under an additive white Gaussian noise (AWGN) channel. Therefore, the theoretical BERs of stream1 and stream2 of our RIS-based 2$\times$2 MIMO-16QAM prototype can be expressed as\footnote{For brevity, we use the approximate BER expression for 16QAM with Gray code under an AWGN channel, which only considers the first and the second terms of the exact expression. More details can be found in \cite{BER}.}
\begin{equation}\label{ss37}
\begin{array}{*{20}{c}}
  {\text{BER}_1} \cong \frac{3}{8}{\text{erfc}}\left({\sqrt{\frac{{{\text{SNR}}_1^{\text{ZF}}}}{{10}}}} \right) + \frac{1}{4}{\text{erfc}}\left( {3\sqrt {\frac{{{\text{SNR}}_1^{\text{ZF}}}}{{10}}} } \right), \\
  {\text{BER}_2} \cong \frac{3}{8}{\text{erfc}}\left({\sqrt{\frac{{{\text{SNR}}_2^{\text{ZF}}}}{{10}}}} \right) + \frac{1}{4}{\text{erfc}}\left( {3\sqrt {\frac{{{\text{SNR}}_2^{\text{ZF}}}}{{10}}} } \right),
\end{array}
\end{equation}
where ${\text{erfc}}\left(\cdot \right)$ is the complementary error function, which is defined as\cite{BER}
\begin{equation}\label{ss38}
{\text{erfc}}\left( x \right) = \frac{2}{{\sqrt \pi  }}\int_x^\infty  {{e^{ - {u^2}}}du}.
\end{equation}
The total BER of the prototype is
\begin{equation}\label{ss39}
{\text{BER}_{\text{total}}} = \frac{{\text{BER}_1} + {\text{BER}_2}}{2}.
\end{equation}
It is worth noting that the above theoretical BERs are obtained under the assumption that the wireless channel is changeless. As explained in footnote4, this assumption can match with the wireless channel condition in the experimental measurement in Section V, so as to properly compare the theoretical BER with the measured BER of the prototype system.

\section{Implementation and Measurement}\label{ImplementationMeasurement}
We present the prototype setup of the RIS-based 2$\times$2 MIMO-QAM wireless system here, which illustrates the hardware architecture, including the detailed hardware modules and their roles in the prototype system. The prototype system realizes real-time RIS-based 2$\times$2 MIMO-QAM transmission over the air. The experimental results demonstrate the feasibility of our proposed method and architecture for realizing RIS-based MIMO-QAM in practice.
\vspace{-0.1cm}
\subsection{Prototype Setup}\label{PrototypeSetup}
\begin{figure*}
\centering
\includegraphics[scale = 0.65]{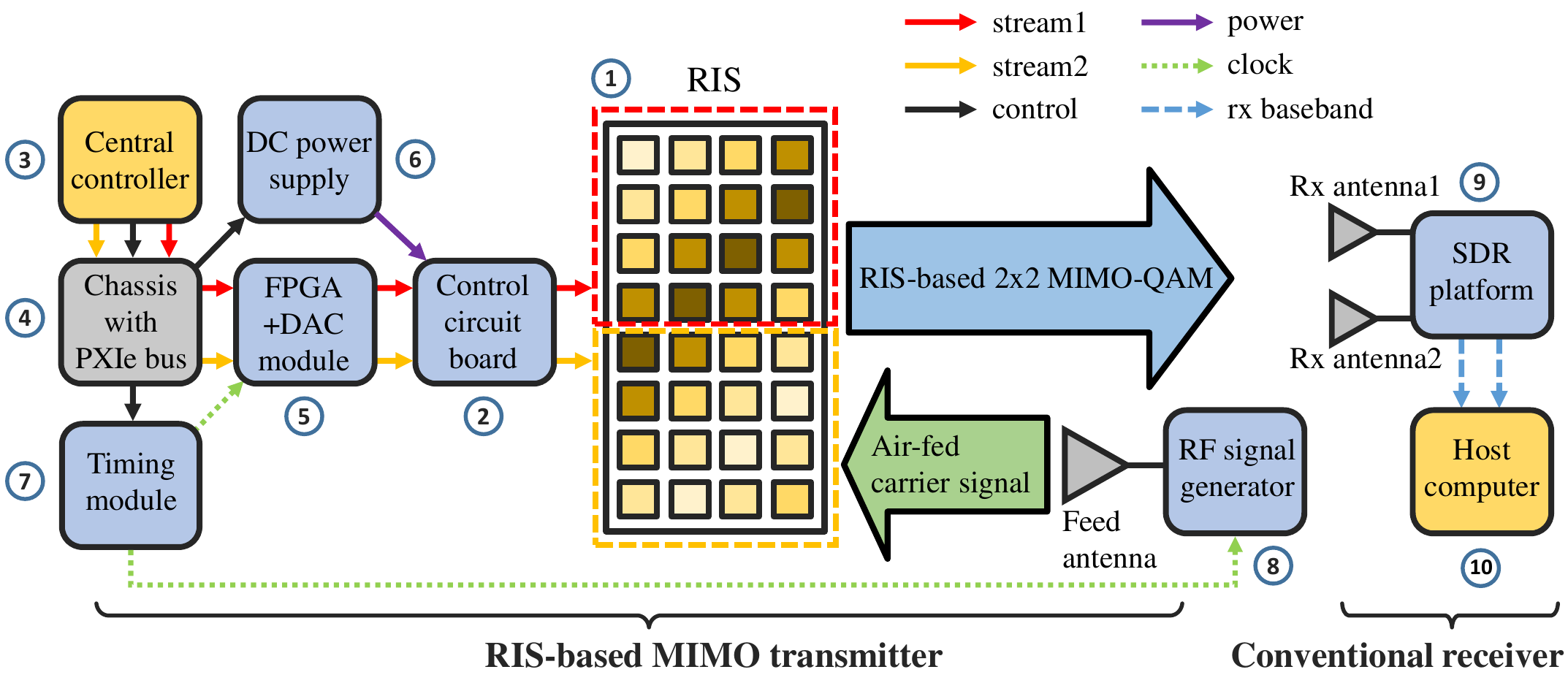}
\caption{The hardware architecture of RIS-based 2$\times$2 MIMO-QAM wireless communication prototype.}
\label{RISArchitecture}
\vspace{-0.2cm}
\end{figure*}

To implement the RIS-based 2$\times$2 MIMO-QAM wireless communication system designed in Section IV, we employ the RIS (programmable metasurface), control circuit board, RF signal generator, several commercial off-the-shelf PXIe modules, software defined radio (SDR) platform, host computer, and antennas as shown in Fig. \ref{RISArchitecture}.

\emph{1) RIS:} The RIS we utilized here is a kind of varactor-diode-based programmable metasurface, which is comprised of  256 unit cells ($N=32$ and $M=8$). The external control voltage signal can change the capacitance of the varactor diodes on its unit cell, thereby controlling the reflection phase. The control voltage signal has an approximate linear relationship with the reflection phase of the unit cell, and can achieve $450^\circ$ continuous phase manipulation range as the control signal varying from $0$V to $21$V. The operating frequency of the RIS is $4.25$ GHz. The simulated gain of each unit cell of RIS is about 9 dBi by assuming 100$\%$ efficiency. The detailed information of the employed RIS can be found in \cite{RIS-based4} and \cite{RIS-based6}.

\emph{2) Control Circuit Board:} The control circuit board is the bridge connecting the DACs and the unit cells of the RIS. The control circuit board has a fixed voltage amplification gain. It amplifies the output voltage signals of the two DACs to respectively control the left half and  right half of the unit cells of the RIS in our prototype system.

\emph{3) Central Controller:} The central controller provides the integrated development environment (IDE) for developing the prototype system. On the central controller, the host program and field programmable gate array (FPGA) program were developed to generate the two source bitstreams, the digital baseband of RIS-based 2$\times$2 MIMO-QAM, and the frame structure. In addition, the central controller performs the necessary control on all the peripheral modules, such as direct current (DC) power supply and timing module.

\emph{4) Chassis with PXIe Bus:} The chassis with PXIe bus acts as the data and control interface between the central controller and the peripheral modules.

\emph{5) FPGA+DAC Module:} There are two 16-bit DACs with $100$ MSps sampling rate (${R_{DAC}}$) inside the FPGA+DAC Module. It converts the digital baseband of RIS-based 2$\times$2 MIMO-QAM transmission with complete frame structure into the two analog voltage signal sequences, which are then delivered as the two input signals of the control circuit board.

\emph{6) DC Power Supply:} The DC power supply provides the $\pm 12$ V  voltage source to the voltage amplifier circuits on the control circuit board.

\emph{7) Timing Module:} The timing module provides the same $10$ MHz reference clock to the FPGA module, the DAC module, and the RF signal generator.

\emph{8) RF Signal Generator:} RF signal generator provides the $4.25$ GHz single-tone RF signal to a horn antenna (feed antenna), which illuminates the RIS as the air-fed carrier signal. The output power of RF signal generator is manually changed during the experiment, so that we can measure the BER performance under different transmission power and SNR condition. It is worth noting that in practical applications, a low-cost single-tone RF signal source can provide this carrier signal instead of the expensive RF signal generator instrument.

\emph{9) SDR Platform:} The SDR platform has two receiving channels that downmix the received RF signals to the baseband signals, and send the digital baseband to the host computer. The two receiving antennas are a kind of single-polarized dipole antenna. The gain of each receiving antenna is 7.4 dBi.

\emph{10) Host Computer:} The host computer processes the digital baseband signals. It performs 2$\times$2 MIMO channel estimation, 2$\times$2 MIMO detection, and QAM demodulation, displays the constellations, and calculates the BER of the recovered bitstreams.

As shown in the left part of Fig. \ref{RISArchitecture}, the RIS-based MIMO transmitter of our prototype system consists of the RIS, the control circuit board, the RF signal generator, and the PXIe system with various peripheral modules. It transmits the RIS-based 2$\times$2 MIMO-QAM wireless frame designed in Section IV. The receiver is depicted in the right part of Fig. \ref{RISArchitecture}, which is composed of the two receiving antennas, the SDR platform, and the host computer. It demodulates the received RIS-based 2$\times$2 MIMO-QAM signals and recovers the transmitted two bitstreams.
\vspace{-0.1cm}
\subsection{Experimental Results}\label{ExperimentalResults}
\begin{figure*}
\centering
\includegraphics[scale = 0.8]{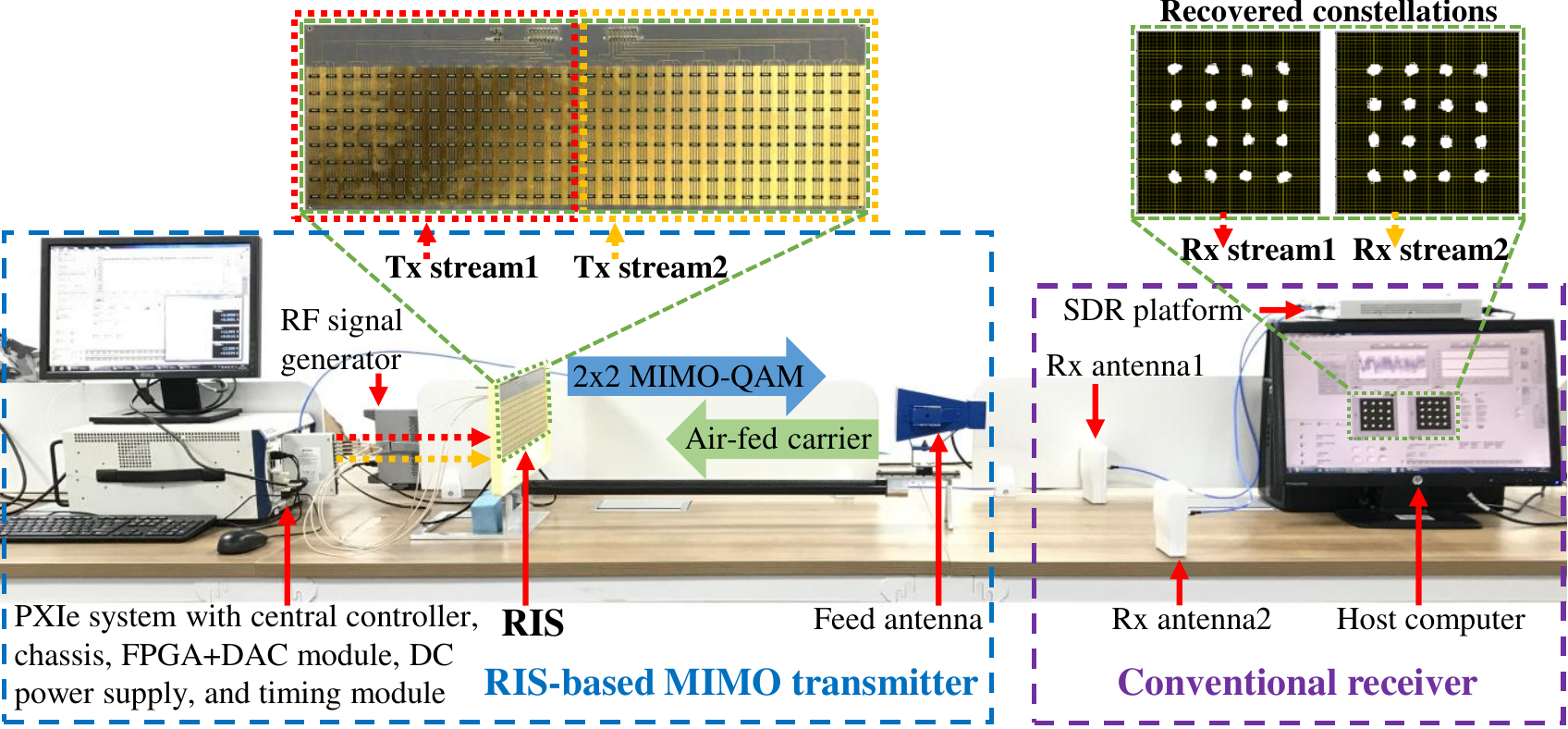}
\caption{A photo of the RIS-based 2$\times$2 MIMO-QAM wireless communication prototype.}
\label{RISPhoto}
\vspace{-0.2cm}
\end{figure*}

The prototype system is shown in Fig. \ref{RISPhoto}. The RIS-based MIMO transmitter is on the left of Fig. \ref{RISPhoto} while the receiver is located on the right. Real-time RIS-based 2$\times$2 MIMO-QAM transmission over the air experiment was conducted indoor. The distance between the RIS and the two receiving antennas is about $1.5$ meters. The recovered constellation diagrams of the two transmitted bitstreams are shown on the upper right hand corner of the figure.

16-QAM modulation and 2$\times$2 MIMO transmission were realized in the prototype. As shown in Fig. \ref{RISPhoto}, the recovered constellation diagrams are clear and dense, which indicates a good BER performance. The data rate of the prototype system reaches $20$ Mbps when ignoring the overhead of the synchronization and pilot subframes. The power consumption of the RIS and the control circuit board is about $0.7$W. The main parameters of the prototype system are summarized in Table \ref{Mainparameters}. As a matter of fact, the transmission rate can be further improved by increasing the size of MIMO, the modulation order, and the symbol rate in the future.

\begin{table}
\centering
\footnotesize
\caption{The main parameters of the prototype.}\label{Mainparameters}
\begin{tabular}{|c|c|}
\hline
\textbf{Parameter} & \textbf{Value}\\
\hline
Carrier Frequency &  $4.25$ GHz\\
\hline
Transmission Scheme &  2$\times$2 MIMO\\
\hline
Modulation Scheme &  16-QAM\\
\hline
Symbol Rate &  $2.5$ MSps\\
\hline
Transmission Rate &  $20$ Mbps\\
\hline
\end{tabular}
\end{table}

\begin{figure}
\centering
\includegraphics[scale = 0.46]{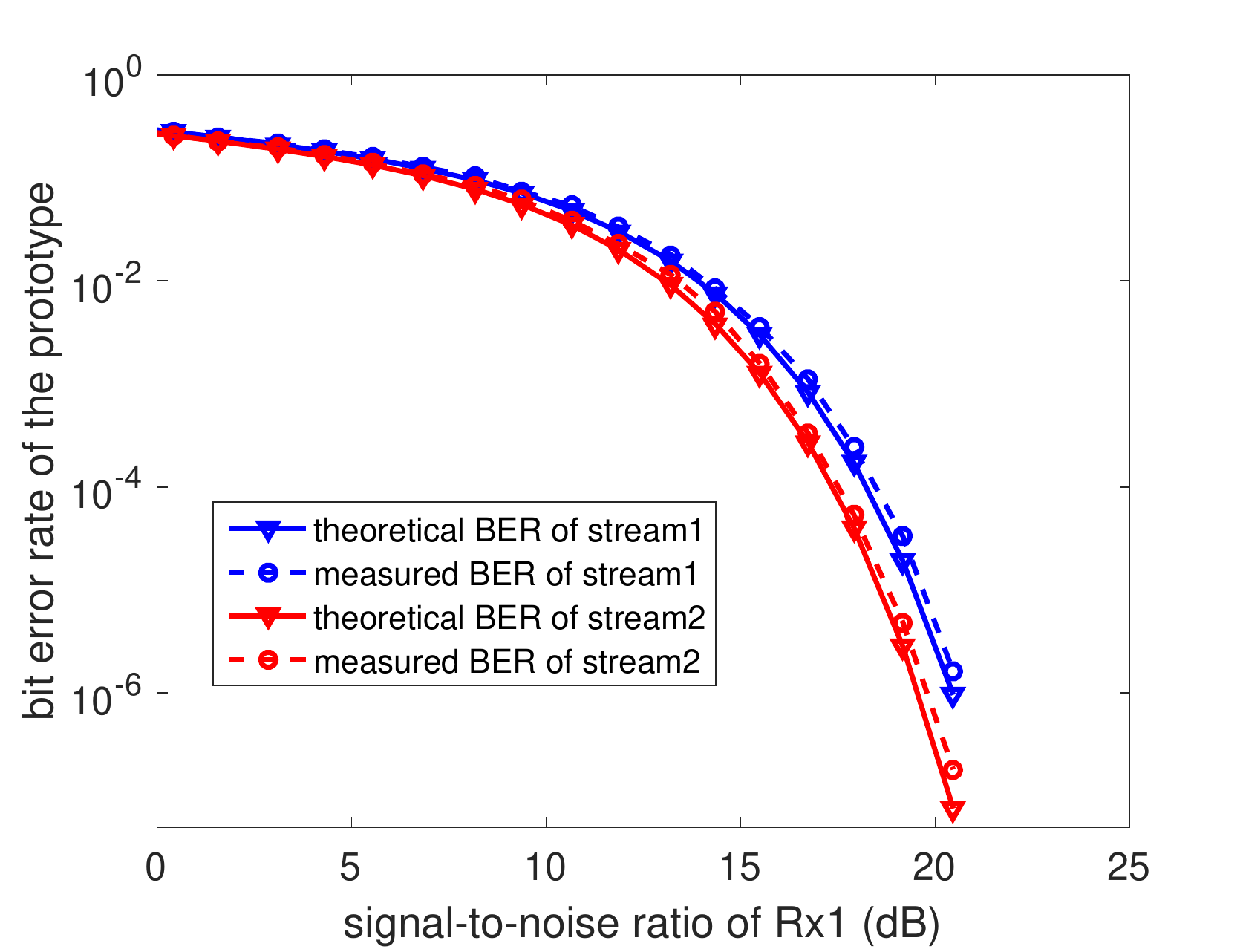}
\caption{The comparison of the measured and theoretical BER performances.}
\label{RISBERnew}
\vspace{-0.2cm}
\end{figure}

We measured the BER performace of the prototype system and compared with its theoretical BER performace. In the BER measurement experiment, $\text {SNR}_1^{\text{ZF}}$ and $\text {SNR}_2^{\text{ZF}}$ described in (\ref{ss36}) can be measured as
\begin{equation}\label{ss40}
\begin{array}{*{20}{l}}
\text {SNR}_1^{\text{ZF}}&= \text{SNR}_{\text{Rx1}}\frac{{\text{SNR}_1^{\text{ZF}}}}{{\text{SNR}_{\text{Rx1}}}}\\
&= \text{SNR}_{\text{Rx1}}\frac{{{{\left| {{{\bar h}_{11}}{{\bar h}_{22}} - {{\bar h}_{12}}{{\bar h}_{21}}} \right|}^2}}}{{\left({{\left| {{{\bar h}_{22}}} \right|}^2} + {{\left| {{{\bar h}_{12}}} \right|}^2}\right)\left({{\left| {{{\bar h}_{11}}} \right|}^2} + {{\left| {{{\bar h}_{12}}} \right|}^2}\right)}}, \\
\text {SNR}_2^{\text{ZF}}&= \text{SNR}_{\text{Rx1}}\frac{{\text{SNR}_2^{\text{ZF}}}}{{\text{SNR}_{\text{Rx1}}}}\\
&= \text{SNR}_{\text{Rx1}}\frac{{{{\left| {{{\bar h}_{11}}{{\bar h}_{22}} - {{\bar h}_{12}}{{\bar h}_{21}}} \right|}^2}}}{{\left({{\left| {{{\bar h}_{21}}} \right|}^2} + {{\left| {{{\bar h}_{11}}} \right|}^2}\right)\left({{\left| {{{\bar h}_{11}}} \right|}^2} + {{\left| {{{\bar h}_{12}}} \right|}^2}\right)}},
\end{array}
\end{equation}
where $\text {SNR}_{\text{Rx1}} = \frac{{p({{\left| {{{\bar h}_{11}}} \right|}^2} + {{\left| {{{\bar h}_{12}}} \right|}^2})}}{{{\sigma ^2}}}$ is the measured SNR of the receiving chain connected to Rx antenna1 as shown in Fig. \ref{RISPhoto}. In the following, we use the measured $\text {SNR}_{\text{Rx1}}$ as the abscissa of the BER curves. By changing the power of the air-fed carrier signal, the BER under different $\text {SNR}_{\text{Rx1}}$ was measured. Meanwhile, the theoretical BER can be calculated according to (\ref{ss37}) and (\ref{ss40}). Fig. \ref{RISBERnew} provides the measured and theoretical BER performances of our prototype. Considering that (\ref{ss37}) is an approximate BER expression as clarified in footnote5, we plot the exact theoretical BER curves in Fig. \ref{RISBERnew}. The measured BER curves of stream1 and stream2 match well with the theoretical ones, which literally validates that the RIS-based transmission has the same BER-SNR performance with the conventional system.

\begin{figure}
\centering
\includegraphics[scale = 0.46]{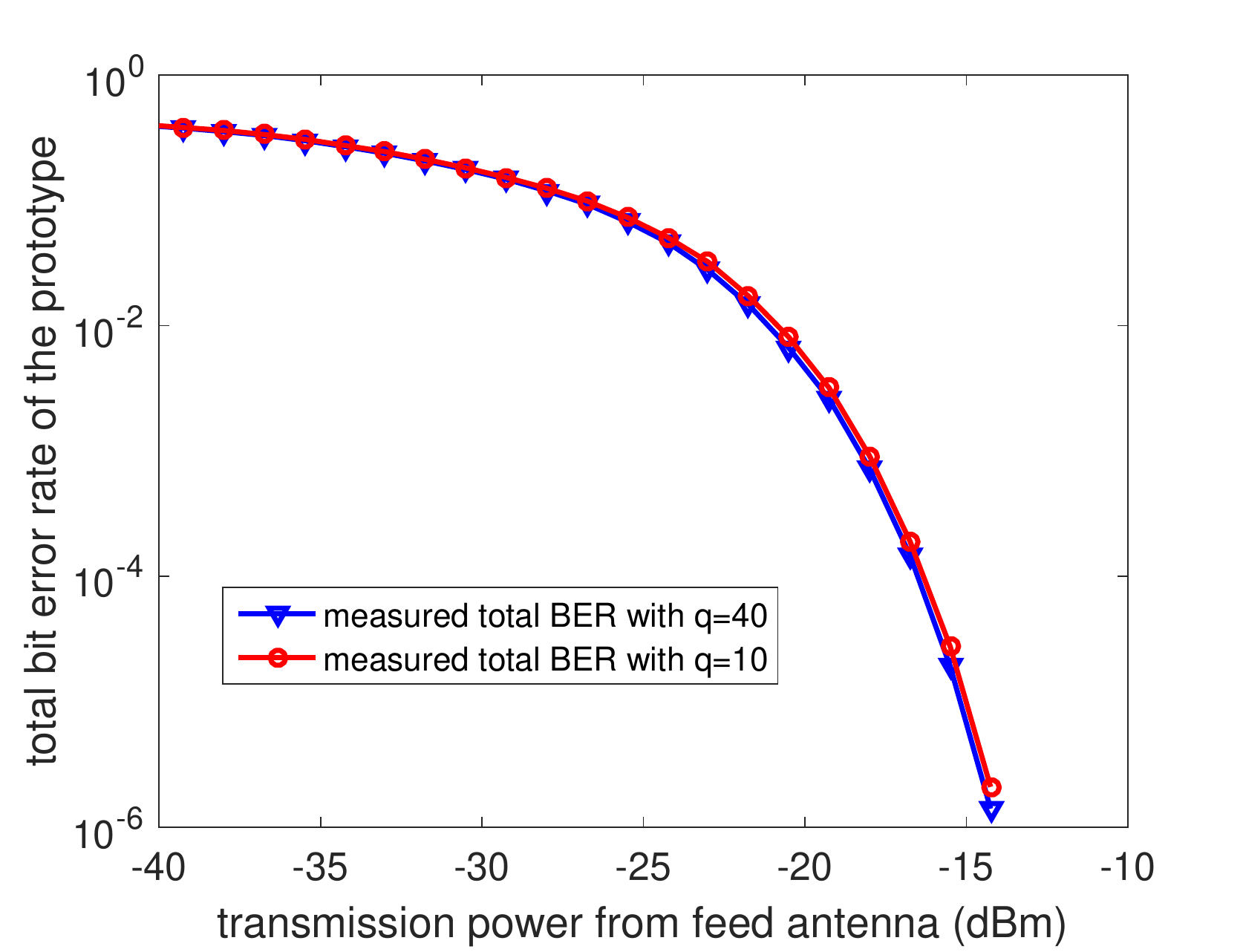}
\caption{The BER under different discrete phase shift steps ($q=40$ and $q=10$) with $2.5$ MSps symbol rate.}
\label{RISBER}
\vspace{-0.2cm}
\end{figure}

In addition, we design a comparative system and conduct experiments to validate the fact that the proposed method of achieving RIS-based QAM is robust with the discrete phase shift steps. Since the maximum sampling rate of the two DACs is $100$ MSps and the symbol rate of the prototype system is $2.5$ MSps, the default number of the discrete phase shift steps in the RIS-based QAM symbol is $40$, i.e., $q=40$. We measured the BER as a function of transmission power under this default condition. To carry out the comparative experiment, we reduced the actual sampling rate of the DACs to $25$ MSps. Then the discrete phase shift steps in the RIS-based QAM symbol became $10$, i.e., $q=10$, and the degree of discretization got worse. The BERs with $q=40$ and $q=10$ were measured and they are provided in Fig. \ref{RISBER}\footnote{We have improved the detection method and redone the experimental measurements. Therefore, the overall BER performance here shown in Fig. 13 is better than that in Fig. 12 of version 1. In version 1, every received data subframe with 64 raw RIS-based QAM symbols is equalized by the pilot subframe with 64 pilot symbols, using a method similar to single carrier frequency domain equalization (do another FFT on raw data subframe for equalization). We found that this method was not proper. The method used now is similar to OFDM with only one effective subcarrier, every raw data on this subcarrier (first order harmonic) is equalized by the average of 64 results of channel estimation in each frame, as described in Section IV. C and D.}. As can be observed, the two BER curves almost coincide with each other, which literally validates that our proposed method is robust with the discrete phase shift steps in the symbol.

\section{Conclusion}\label{Conclusion}
In this paper, we have presented the analytical formulation describing the system model of RIS-based MIMO-QAM wireless communications considering the physics and EM nature of the RISs. A basic method and the transceiver design of achieving high-order modulation such as QAM and MIMO transmission through the RISs have been presented. In addition, the hardware constraints of the RISs including phase dependent amplitude and discrete phase shift, and their impacts on the system design were discussed and analyzed. Moreover, we have presented the details of our prototype that had been implemented to realize real-time RIS-based 2$\times$2 MIMO 16QAM transmission over the air with $20$ Mbps data rate, and our experimental results have validated convincingly that the proposed RIS-based MIMO-QAM transmitter architecture is robust. These encouraging results suggest that RISs provide an attractive architecture for realizing UM-MIMO and holographic MIMO technologies, with affordable hardware complexity.

\begin{appendices}
\section{}\label{A}
\centerline{Proof of Theorem 1}
The power of the incident single-tone carrier signal into unit cell $U_{n,m}$ can be expressed as
\begin{equation}\label{sss1}
P_{n,m}^{in} = S{d_x}{d_y},
\end{equation}
and the electric field of the incident single-tone carrier signal into $U_{n,m}$ is given by \cite{Book0}
\begin{equation}\label{sss2}
E_{n,m}^{in} = \sqrt {2{Z_0}S}{e^{j2\pi {f_c}t}},
\end{equation}
where $Z_0$ is the characteristic impedance of the air, and $f_c$ is the frequency of the incident single-tone carrier signal. According to the law of energy conservation, for the unit cell $U_{n,m}$, the power of the incident signal times the square of the reflection coefficient is equal to the total power of the reflected signal. As such, we have
\begin{equation}\label{sss3}
P_{n,m}^{in}{|}\Gamma _{n,m}^2{|} = P_{n,m}^{reflect},
\end{equation}
where $P_{n,m}^{reflect}$ denotes the total reflected signal power of the unit cell $U_{n,m}$, and the reflection coefficient is represented as ${\varGamma _{n,m}} = {A_{n,m}}{e^{j{\varphi _{n,m}}}}$ as originally defined in (\ref{ss1}). Also, the power of the reflected signal received by the $k^{th}$ receiving antenna from $U_{n,m}$ can be expressed as
\begin{equation}\label{sss4}
\begin{aligned}
&P_{n,m}^{k} =\\
&\frac{{{G}{P_{n,m}^{reflect}}}}{{4\pi d{{_{n,m}^k}^2}}}F\left( {\theta _{n,m}^{AOD,k},\phi _{n,m}^{AOD,k}} \right){F^{rx}}\left( {\theta _{n,m}^{AOA,k},\phi _{n,m}^{AOA,k}} \right)A_r,
\end{aligned}
\end{equation}
where $A_r$ represents the aperture of the receiving antenna.

By combining (\ref{sss1}), (\ref{sss3}) and (\ref{sss4}), the electric field of the reflected signal received by the $k^{th}$ receiving antenna from $U_{n,m}$ is obtained as (\ref{sss5}) shown at the top of the next page \cite{Book0},
\begin{figure*}[!t]
\begin{align}\label{sss5}
E_{n,m}^k &= \sqrt {\frac{{2{Z_0}P_{n,m}^k}}{{{A_r}}}} {e^{j(\frac{{ - 2\pi d_{n,m}^k}}{\lambda }+{\varphi _{n,m}})}}{e^{j2\pi {f_c}t}}\notag\\
&= \sqrt {\frac{{2{Z_0}GS{d_x}{d_y}F\left( {\theta _{n,m}^{AOD,k},\phi _{n,m}^{AOD,k}} \right){F^{rx}}\left( {\theta _{n,m}^{AOA,k},\phi _{n,m}^{AOA,k}} \right)}}{{4\pi d{{_{n,m}^k}^2}}}} {\Gamma _{n,m}}{e^{\frac{{ - j2\pi d_{n,m}^k}}{\lambda }}}{e^{j2\pi {f_c}t}}\notag\\
&= \frac{{\sqrt {{Z_0}GF\left( {\theta _{n,m}^{AOD,k},\phi _{n,m}^{AOD,k}} \right){F^{rx}}\left( {\theta _{n,m}^{AOA,k},\phi _{n,m}^{AOA,k}} \right)} }}{{\sqrt {2\pi } d_{n,m}^k}}{e^{\frac{{ - j2\pi d_{n,m}^k}}{\lambda }}}\sqrt {S{d_x}{d_y}} {A_{n,m}}{e^{j{\varphi _{n,m}}}}{e^{j2\pi {f_c}t}}
\end{align}
\hrulefill
\end{figure*}
where $(\frac{{ - 2\pi d_{n,m}^k}}{\lambda }+{\varphi _{n,m}})$ is the phase alteration caused by the propagation and the reflection coefficient of $U_{n,m}$. The total electric field of the received signal of the $k^{th}$ receiving antenna is the superposition of the electric fields reflected by all the unit cells towards it, which can be written as
\begin{equation}\label{sss6}
{E_k} = \sum\limits_{m = 1}^{M} {\sum\limits_{n = 1}^{N} {E_{n,m}^k} }.
\end{equation}
The instantaneous received signal power of the $k^{th}$ receiving antenna can be obtained as
\begin{equation}\label{sss7}
{P_k} = \frac{{E_k^2}}{{2{Z_0}}}{A_r},
\end{equation}
where the aperture of the $k^{th}$ receiving antenna can be written as
\begin{equation}\label{sss8}
{A_r} = \frac{{{G_r}{\lambda ^2}}}{{4\pi }}.
\end{equation}

By substituting (\ref{sss5}), (\ref{sss6}), and (\ref{sss8}) into (\ref{sss7}), the instantaneous received signal power of the $k^{th}$ receiving antenna is obtained as (\ref{sss9}) shown at the top of the next page.
\begin{figure*}[!t]
\begin{align}\label{sss9}
{P_k} = \frac{{G{G_r}{\lambda ^2}S{d_x}{d_y}}}{{16{\pi ^2}}}{\left( {\sum\limits_{m = 1}^M {\sum\limits_{n = 1}^N {\frac{{\sqrt {F\left( {\theta _{n,m}^{AOD,k},\phi _{n,m}^{AOD,k}} \right){F^{rx}}\left( {\theta _{n,m}^{AOA,k},\phi _{n,m}^{AOA,k}} \right)} }}{{d_{n,m}^k}}{e^{\frac{{ - j2\pi d_{n,m}^k}}{\lambda }}}{\Gamma _{n,m}}{e^{j2\pi {f_c}t}}} } } \right)^2}
\end{align}
\end{figure*}
As the received signal can be expressed by the square root of the instantaneous received signal power, we therefore get (\ref{sss10}) shown at the top of the next page.
\begin{figure*}[!t]
\begin{align}\label{sss10}
{y_k} = \sum\limits_{m = 1}^M {\sum\limits_{n = 1}^N {\frac{{\sqrt {G{G_r}{\lambda ^2}F\left( {\theta _{n,m}^{AOD,k},\phi _{n,m}^{AOD,k}} \right){F^{rx}}\left( {\theta _{n,m}^{AOA,k},\phi _{n,m}^{AOA,k}} \right)S{d_x}{d_y}} }}{{4\pi d_{n,m}^k}}{e^{\frac{{ - j2\pi d_{n,m}^k}}{\lambda }}}{\Gamma _{n,m}}{e^{j2\pi {f_c}t}}} }
\end{align}
\hrulefill
\end{figure*}

\section{}\label{B}
\centerline{Proof of Theorem 2}

Given the periodic signal
\begin{equation}\label{sss11}
{s}(t) = {e^{j \frac{{\Delta \varphi }}{{{T_s}}}t}}, ~\mbox{for }t \in \left[ {0,{T_s}} \right],
\end{equation}
its exponential Fourier series expansion is given by
\begin{equation}\label{sss12}
s(t) = \sum\limits_{l =  - \infty }^\infty  {{a_l}{e^{jl\frac{{2\pi }}{{{T_s}}}t}}},
\end{equation}
where
\begin{equation}\label{sss13}
\begin{aligned}
{a_l} &= \frac{1}{{{T_s}}}\int_0^{{T_s}} {s(t){e^{ - jl\frac{{2\pi }}{{{T_s}}}t}}dt} =  \frac{1}{{{T_s}}}\int_0^{{T_s}} {{e^{j\frac{{\Delta \varphi }}{{{T_s}}}t}}{e^{ - jl\frac{{2\pi }}{{{T_s}}}t}}dt}\\
&=  \frac{{1}}{{{T_s}}}\int_0^{{T_s}} {{e^{j\left(\frac{{\Delta \varphi }}{{{T_s}}} - l\frac{{2\pi }}{{{T_s}}}\right)t}}dt}.
\end{aligned}
\end{equation}
If $\Delta \varphi  - 2l\pi  = 0$, (\ref{sss13}) can be further expressed as
\begin{equation}\label{sss14}
{a_l} = \frac{{1}}{{{T_s}}}\int_0^{{T_s}} {e^{j0}}dt = 1.
\end{equation}
On the contrary, if $\Delta \varphi  - 2l\pi  \ne 0$, (\ref{sss13}) can be further expressed as
\begin{equation}\label{sss15}
\begin{aligned}
{a_l} =& \frac{{1}}{{{T_s}}}\int_0^{{T_s}} {{e^{j\left(\frac{{\Delta \varphi }}{{{T_s}}} - l\frac{{2\pi }}{{{T_s}}}\right)t}}dt}\\
=& \frac{{1}}{{{T_s}j\left(\frac{{\Delta \varphi }}{{{T_s}}} - l\frac{{2\pi }}{{{T_s}}}\right)}}\left[{e^{j\left(\frac{{\Delta \varphi }}{{{T_s}}} - l\frac{{2\pi }}{{{T_s}}}\right){T_s}}} - 1\right]\\
=& \frac{{j(1 - {e^{j(\Delta \varphi  - 2l\pi )}})}}{{\Delta \varphi  - 2l\pi }}\\
=& \frac{{1}}{{\frac{{\Delta \varphi }}{2} - l\pi }}\left[j\frac{{1 - \cos (\Delta \varphi  - 2l\pi )}}{2} + \frac{{\sin (\Delta \varphi  - 2l\pi )}}{2}\right]\\
=& \frac{{1}}{{\frac{{\Delta \varphi }}{2} - l\pi }}\left[j{\sin ^2}\left(\frac{{\Delta \varphi }}{2} - l\pi \right)\right.\\
&+\left. \sin \left(\frac{{\Delta \varphi }}{2} - l\pi \right)\cos \left(\frac{{\Delta \varphi }}{2} - l\pi \right)\right]\\
=& \frac{{\sin \left(\frac{{\Delta \varphi }}{2} - l\pi \right)}}{{\frac{{\Delta \varphi }}{2} - l\pi }}\left[j\sin \left(\frac{{\Delta \varphi }}{2} - l\pi \right) + \cos \left(\frac{{\Delta \varphi }}{2} - l\pi \right)\right]\\
=& \operatorname{sinc} \left(\frac{{\Delta \varphi }}{2} - l\pi \right){e^{j\left(\frac{{\Delta \varphi }}{2} - l\pi \right)}}\\
=& \left| {\operatorname{sinc} \left(\frac{{\Delta \varphi }}{2} - l\pi \right)} \right|\\
&\times {e^{j\left( {\frac{{\Delta \varphi }}{2} - l\pi  + \bmod \left(\left\lfloor {\frac{{\Delta \varphi }}{{2\pi }} - l} \right\rfloor,2\right) \cdot \pi  + \varepsilon (2l\pi  - \Delta \varphi ) \cdot \pi } \right)}}.
\end{aligned}
\end{equation}
By combining (\ref{sss14}) and (\ref{sss15}), we can get
\begin{equation}\label{sss16}
\begin{aligned}
{a_l} =& \left| {\operatorname{sinc} \left(\frac{{\Delta \varphi }}{2} - l\pi \right)} \right|\\
&\times {e^{j\left( {\frac{{\Delta \varphi }}{2} - l\pi  + \bmod \left(\left\lfloor {\frac{{\Delta \varphi }}{{2\pi }} - l} \right\rfloor,2\right) \cdot \pi  + \varepsilon (2l\pi  - \Delta \varphi ) \cdot \pi } \right)}}.
\end{aligned}
\end{equation}
Note that the baseband symbol defined in Theorem 2 has a circular time shift $t_0$ compared with the signal defined in (\ref{sss11}). According to the time delay property of the Fourier transform, Theorem 2 can be obtained, which completes the proof.
\end{appendices}

\end{document}